\documentclass[journal]{IEEEtran}
\usepackage{xcolor}
\usepackage{graphicx}
\usepackage{booktabs}
\usepackage[table]{xcolor}
\usepackage{qcircuit}
\usepackage{acronym}
\usepackage{amsfonts}
\usepackage{braket}
\usepackage{subcaption}

\usepackage{cite}
\usepackage[cmex10]{amsmath}
\usepackage{url}

\acrodef{ML}[ML]{machine learning}
\acrodef{QK}[QK]{quantum kernel}
\acrodef{QML}[QML]{quantum machine learning}
\acrodef{SVM}[SVM]{support vector machine}
\acrodef{ML}[ML]{machine learning}

\newcommand{\Artur}[1]{\textcolor{blue}{#1}}
\newcommand{\Amer}[1]{{\textcolor{red}{#1}}}

\begin{document}
\title{Large-Scale Quantum Kernels for Hyperspectral Data Classification}
\author{Amer~Delilbasic,~\IEEEmembership{Student Member,~IEEE,}
        Artur~Miroszewski,~\IEEEmembership{Member,~IEEE,}
        Agata~Wijata,~\IEEEmembership{Member,~IEEE,}
        Jakub~Nalepa,~\IEEEmembership{Senior Member,~IEEE,}
        Jakub~Mielczarek, Morris~Riedel, and~Gabriele~Cavallaro, \IEEEmembership{Senior~Member,~IEEE}
\thanks{A. Delilbasic, M. Riedel and G. Cavallaro are with Forschungszentrum J\"ulich, 52428 J\"ulich, Germany, and University of Iceland, 102 Reykjavík, Iceland. E-mail:\{a.delilbasic, m.riedel, g.cavallaro\}@fz-juelich.de.}
\thanks{A. Miroszewski and J. Mielczarek are with Jagiellonian University, Łojasiewicza 11, 30-348 Krakow, Poland. E-mail: \{artur.miroszewski, jakub.mielczarek\}@uj.edu.pl.}
\thanks{A. Wijata and J. Nalepa are with Silesian University of Technology, Akademicka 2A, 44-100 Gliwice, Poland, and KP Labs, Bojkowska 37J, 44-100, Gliwice, Poland. E-mail: \{awijata, jnalepa\}@ieee.org.}
\thanks{Manuscript received August 29, 2025; revised August 30, 2025.}}

%
%

\markboth{IEEE Journal of Selected Topics in Applied Earth Observations and Remote Sensing, 2025}%
{Delilbasic \MakeLowercase{\textit{et al.}}: Bare Demo of IEEEtran.cls for Journals}
%



\maketitle

\begin{abstract}

Quantum kernel methods have emerged as a promising approach for leveraging high-dimensional feature spaces in machine learning, particularly in domains where classical kernel methods face scalability limitations. In this work, we present the first large-scale study of fidelity-quantum-kernel support vector machines for hyperspectral data classification without requiring heavy prior feature selection or dimensionality reduction. By simulating quantum kernels using tensor network contraction techniques and GPU acceleration, we overcome the computational bottlenecks traditionally associated with quantum models, achieving quadratic scaling $O(n^2)$ in the number of qubits. Our approach enables the evaluation of quantum kernels on hyperspectral data with hundreds of spectral bands, aligning quantum feature spaces with real-world remote sensing applications. We provide an in-depth analysis of kernel bandwidth optimization, demonstrating its crucial role in mitigating exponential concentration effects and ensuring the model's ability to generalize. Experimental results on binary classification (Indian Pines and Methane Detection) and multiclass classification (Indian Pines) demonstrate that quantum kernels achieve competitive performance compared to a broad range of state-of-the-art classical baselines. As illustrative cases, on four 50-band splits selected from Indian Pines, the quantum model achieved a $78.0 \pm 6.2$\% accuracy for a binary classification task compared to $72.0 \pm 5.0$\% for the standard radial basis function (RBF) kernel. For a four-class classification task, the quantum kernel reached $83.3 \pm 3.1$\% accuracy, outperforming several state-of-the-art baselines. On five 75-band splits selected from the Methane Detection dataset, the quantum approach yielded $58.5 \pm 5.0$\% accuracy versus $55.1 \pm 2.5$\% for the classical counterpart. This study provides a foundation for future exploration of quantum machine learning in high-dimensional Earth observation tasks.

\end{abstract}


%
\IEEEpeerreviewmaketitle

\section{Introduction}

\IEEEPARstart{H}{yperspectral} remote sensing has become a cornerstone of modern Earth observation (EO), offering unparalleled spectral resolution across hundreds of contiguous bands~\cite{2023_Wijata}. Unlike traditional multispectral sensors, which capture only a few broad bands, hyperspectral instruments provide fine-grained spectral signatures of surface materials, enabling accurate identification and classification of vegetation, minerals, water bodies, and urban features~\cite{bioucas2013hyperspectral,2021_Nalepa}. These rich, high-dimensional datasets open new possibilities for environmental monitoring and resource management, but they also pose substantial computational challenges.

\Ac{QML}~\cite{schuld2015introduction,biamonte2017quantum} has emerged as a promising research area at the intersection of quantum computing and data-driven modeling. \ac{QML} algorithms employ quantum circuits that use quantum mechanical effects to encode and process information. An important expectation is that well-designed \ac{QML} models should be difficult to simulate on classical hardware, at the same time offering strong learning performance that can eventually be realized on near-term quantum devices. Thus, \ac{QML} algorithms may serve as an extension of the \ac{ML} practitioner’s toolbox, providing complementary models that are not accessible without quantum computers.

Among candidate approaches, quantum kernel methods, particularly quantum kernel support vector machines~\cite{schuld2019quantum}, stand out as a leading direction. These methods embed classical data into quantum states, granting access to exponentially large feature spaces. In principle, this property could enhance the separability of complex datasets and yield advantages over classical kernel methods~\cite{havlivcek2019supervised}. Until recently, a major obstacle was that studying the behavior of large-scale quantum kernels was considered computationally infeasible, both on real, noisy hardware or using simulators.

We are now in a particularly interesting situation. Recent advances by Chen et al.~\cite{chen2025validating} demonstrate that fidelity quantum kernels~\cite{havlivcek2019supervised} can be efficiently simulated at scales previously deemed out of reach, thanks to tensor-network techniques and graphics processing units (GPUs) acceleration. This breakthrough provides a unique opportunity: for the first time, the behavior of quantum fidelity kernels defined on large quantum devices (on the order of hundreds of qubits) can be systematically analyzed. This enables an empirical evaluation of the widely held claim that the power of \ac{QML} stems from the expressivity of high-dimensional feature spaces. This work provides the initial results for \ac{QML} models at this scale, together with observations, practical recommendations, and a technical pathway for the future processing of larger and higher-dimensional remote sensing data.
At the same time, the fact that these kernels can be simulated classically undermines their potential to deliver a robust quantum advantage. Importantly, this limitation applies only to the family of fidelity kernels with low-entangling circuits; other quantum kernel constructions, such as projected quantum kernels~\cite{huangPowerDataQuantum2021}, remain classically intractable and continue to represent promising candidates for future quantum advantage. 

This development is especially relevant for hyperspectral imaging, where the dimensionality of data (often hundreds of spectral bands) matches the scales at which these new simulations are tractable. By analyzing hyperspectral datasets without feature selection or dimensionality reduction, one can avoid the criticism that pre-processing may trivialize the classification task and mask the intrinsic capabilities of quantum models. In this work, we follow this principle and, to the best of our knowledge, present the first comprehensive study of quantum support vector machines applied to unmodified (with respect to the number of available spectral bands) hyperspectral datasets.

\subsection{Related Work}

The application of quantum machine learning (\ac{QML}) methods to EO data is an active field of research. For recent reviews, see~\cite{torta2025quantum, quovadis, sebastianelli2025quantum}. 
In particular, different \ac{QML} methods have been applied to the analysis of multispectral~\cite{henderson2021methods, miroszewski2023detecting, gawron2020multi, sebastianelli2021circuit, sebastianelli2025quanv4eo}, hyperspectral~\cite{lin2024quantum, priyanka2024hyperspectral, shaik2022quantum, shaik2022accuracy, otgonbaatar2021quantum, wang2026queennet}, and synthetic-aperture radar (SAR) data~\cite{mauro2024qspecklefilter, miller2023quantum, otgonbaatar2021natural, bhattacharya2025bloch}. However, due to the limitations of current quantum devices in the noisy intermediate-scale quantum (NISQ) era~\cite{preskillQuantumComputingNISQ2018}, most of the work in this area is carried out on classical simulators of quantum computers~\cite{bergholm2018pennylane, javadi2024quantum}. This, in turn, requires the use of feature and sample reduction techniques, such as principal component analysis~\cite{priyanka2024hyperspectral, shaik2022quantum}, superpixel segmentation~\cite{miroszewski2023detecting, wijata2024detection}, or coreset~\cite{otgonbaatar2021assembly} methods. This issue is particularly pronounced in hyperspectral data, where the high number of spectral bands makes dimensionality reduction almost unavoidable. Some studies have also employed quantum annealers, taking advantage of their relatively higher technical fidelity~\cite{delilbasic2021quantum, otgonbaatar2021assembly, delilbasic2023single, pasetto2022quantum}. Nevertheless, because of data encoding overheads and the limited capabilities of quantum annealers, these works also predominantly rely on reduced datasets. To the best of our knowledge, the use of full spectral information in large-scale simulations of \ac{QML} models for EO tasks remains unexplored, and this is the gap that we address in this work.

\subsection{Contributions}
The contributions of this work are organized into the primary and secondary ones to clearly distinguish the core advances from supporting analyses.

\textbf{Main contributions:}
\begin{itemize}
    \item \textit{Highly-dimensional quantum kernel simulations for hyperspectral data}:
We present, to the best of our knowledge, the first study of quantum kernel methods applied to hyperspectral Earth observation datasets using the \emph{full spectral range}, without feature selection or dimensionality reduction. By leveraging tensor-network-based simulations, we enable quantum kernel evaluations in regimes involving hundreds of input features and many-qubit systems, which were previously inaccessible to classical simulation approaches.
    \item \textit{Demonstration of strong performance in classification and detection tasks}:  
We demonstrate that high-qubit-number quantum kernel SVMs achieve strong performance in both multi-class and binary classification tasks on hyperspectral datasets (concerning land-cover mapping and methane detection, respectively).
\end{itemize}
In support of the main contributions, we provide the following additional analyses.

\textbf{Secondary contributions:}
\begin{itemize}
    \item \textit{Benchmarking tensor-network-based quantum kernel simulation}:
We compare the tensor-network-based \texttt{cuTN-QSVM} library~\cite{chen2025validating} with standard \texttt{statevector} simulation in terms of computational cost, highlighting the scalability advantages that enable high-dimensional quantum kernel studies.
    \item \textit{Analysis of quantum kernel behavior in many-qubit regimes}:  
Using tools from classical machine learning, quantum machine learning, and spectral kernel theory, we provide an in-depth characterization of high-qubit-number quantum kernels.
This analysis clarifies how increasing system size impacts the geometry of the induced feature space and the resulting learning performance.
We argue for the necessity of restricting the effective set of quantum states in high-dimensional settings and show that kernel bandwidth optimization can play a crucial role in that. We identify geometric and inductive-bias differences that emerge after bandwidth optimization.
\end{itemize}

\subsection{Article Structure}

The remainder of this paper is organized as follows:

\begin{itemize}
    
    \item Section~II provides theoretical background on classical and quantum kernel methods, with a focus on the fidelity quantum kernel and the role of kernel bandwidth in mitigating expressibility and concentration effects.
    
    \item Section~III presents the classification pipeline for hyperspectral data using quantum kernels, including pixel-wise support vector machines (SVMs) formulation and quantum circuit simulation.
    
    \item Section~IV describes the experimental setup, including datasets (Indian Pines and Methane Detection), model training via Bayesian optimization, and implementation using tensor network simulation.
    
    \item Section~V reports the shown metrics and their significance for model performance and kernel analysis.

    \item Section~VI discusses the results in detail, including runtime comparisons, bandwidth optimization trends, classification accuracy, kernel expressibility, alignment, geometric difference, and spectral analysis.
    
    \item Section~VII concludes the paper and discusses implications for future quantum machine learning applications in remote sensing.
    
\end{itemize}

\section{Theoretical Background}
\label{sec:background}
\subsection{Kernel Methods}

Kernel methods in \ac{ML} are a class of algorithms that are characterized by mapping data into a high-dimensional (or even infinite-dimensional) feature space, where the inner products between transformed data points are computed. This allows complex patterns to be learned with relatively simple models~\cite{hofmann2008kernel}, including notable representatives of \ac{ML} methods such as SVMs, principal component analysis, and ridge regression~\cite{bishop2006pattern}.
The mapping of data belonging to some initial space $x \in \chi$ to a target feature space $\mathcal{H}$, $\phi(x): \chi \mapsto\mathcal{H}$ may involve significant computational effort. However, according to Mercer’s theorem, any continuous, symmetric, positive semi-definite function $K(x,y)$, defined on a compact domain, can be written as an inner product in a (possibly infinite-dimensional) feature space:
\begin{equation}
\label{eq:kernel-function}
K(x,y) = \langle \phi(x), \phi(y) \rangle_\mathcal{H}.    
\end{equation}
Due to this equivalence, in many cases we do not need to explicitly compute the costly feature transformations $\phi(x), \phi(y)$ and their inner product. Instead, we can directly evaluate the \textit{kernel} function $K(x,y)$, which often has a much simpler form. This approach is known as the \textit{kernel trick} and is one of the key reasons why the classical kernel methods became so effective and powerful.
A notable example of the use of the kernel trick is the kernel function family of RBFs~\cite{scholkopf2002learning}. The RBF kernel function is simply a Gaussian centered at one point and evaluated at the second point. With explicit computation of the transformations of data points, we would have to operate on infinite-dimensional spaces. Moreover, the use of the kernel trick in this case allows for a clear interpretation of the similarity measure between data points.
Kernel methods are valued for their mathematical elegance, with strong theoretical guarantees, as well as their versatility and interpretability~\cite{hofmann2008kernel, shawe2004kernel}. However, with at least quadratic scaling of the computational complexity in the number of data points, kernel methods have recently been superseded by neural networks in many cases, especially in big data applications~\cite{sra2011optimization}.
Nevertheless, there are still active domains of research in kernel methods, one of them being quantum kernels~\cite{quovadis}.

\begin{figure*}[ht]
    \begin{subfigure}[b]{0.35\linewidth}
        \centering
        \includegraphics[width=\linewidth]{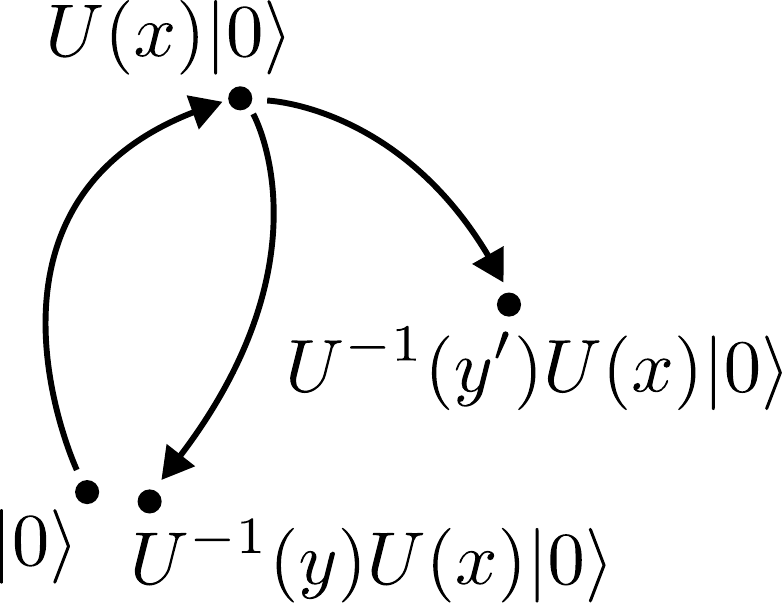}
        \caption{}
    \end{subfigure}
    \begin{subfigure}[b]{0.65\linewidth}
        \centering
        \includegraphics[width=\linewidth]{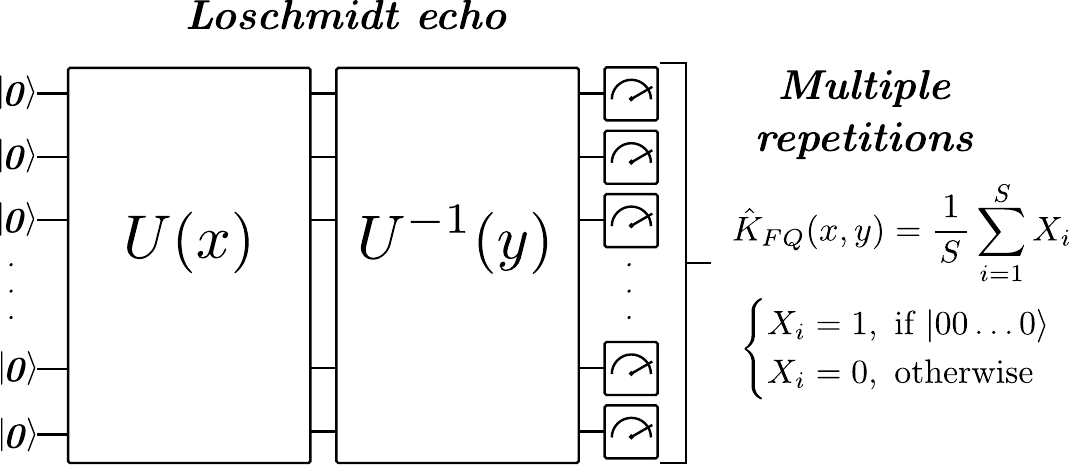}
        \caption{}
    \end{subfigure}

    \caption{The Loschmidt echo approach to fidelity quantum kernel estimation. One starts with the initial state $|0\rangle$, implements a unitary transformation parameterized by the first data point $U(x)$, and then applies the subsequent inverse unitary transformation parameterized by a second data point. 
    (a) If the second data point is similar to the first one, the combined transformation-inverse transformation $U^{-1}(y)U(x)$ should result in a state close to the initial $|0\rangle$ state. Conversely, if the second data point is not similar to $x$, then the total transformation $U^{-1}(y')U(x)$ takes the state far from the initial state $|0\rangle$.
    (b) The Loschmidt echo circuit is run $S$ times in order to gather statistics for the estimation of the fidelity quantum kernel value.}
    \label{fig:FQK_idea}
\end{figure*}

\subsection{Quantum Kernel Methods}\label{sec:QKMs}

The general idea of quantum kernel methods 
is to expand the set of available kernel functions by computing their values on a quantum computer. In this case, we do not rely on the classical kernel trick; instead, we explicitly construct feature maps and use quantum routines to compute the corresponding inner products.
The first advantage of this approach is the relatively cheap access to high-dimensional feature spaces. In principle, the dimensionality of these spaces grows exponentially, as $\mathcal{O}(2^n)$, with the size of the quantum device, where $n$ denotes the number of qubits. Moreover, the processing of quantum information is based on unitary operations and is fundamentally different from the processing of classical bits. For this reason, the set of functions naturally generated by a quantum computer is vastly different from the functions typically used in classical computing. As a result, quantum kernel methods provide access to a large set of novel and yet unexplored similarity functions.

The original quantum kernel family, introduced in \cite{havlivcek2019supervised, schuld2019quantum}, is the fidelity quantum kernel, defined as
\begin{equation}\label{eq:fqk}
    K_{FQ}(x,y) = | \langle \phi(y) | \phi(x) \rangle |^2,
\end{equation}
where $| \phi(z) \rangle$ is the quantum state parameterized with the data point $z$.
Parameterization is performed by applying a $z$-parameterized unitary transformation, $U(z)$, to some (usually initial) quantum state, $|\phi(z)\rangle = U(z)|e\rangle$. This process is called quantum data embedding or quantum data encoding. It is easy to see from the above definition that indeed the fidelity quantum kernel is positive-definite and symmetric. 
From a practical perspective, one can obtain the fidelity quantum kernel on a quantum device in many ways. 
The most popular are the SWAP test~\cite{barenco1997stabilization, buhrman2001quantum} and the Loschmidt echo~\cite{peres1984stability} approach. We focus on the latter because it requires only $n$ qubits instead of $2n+1$, significantly reducing the computational resources needed for simulation.
In the equivalent formulation of density matrices, Eq.~(\ref{eq:fqk}) can be written as
\begin{align}\label{eq:fqk_density}
    K_{FQ}(x,y) &= \text{Tr} \left[ \rho(y) \rho(x) \right] \nonumber \\
        &= \text{Tr} \left[ U(y) \rho_0 U^{\dagger}(y) U(x) \rho_0 U^{\dagger}(x)\right]\\
        &= \text{Tr}\left[ \rho_0 \{U^{\dagger}(y) U(x) \rho_0 U^{\dagger}(x)U(y)\}\right], \nonumber
\end{align}
where $\rho_0 = |e\rangle \langle e|$ is the initial state density matrix, and in the last line we used the cyclic property of the trace. From this point on, we will assume that the initial state is the vacuum state in the computational basis $|e\rangle = |0\rangle$. From the last line of Eq.~(\ref{eq:fqk_density}) we can see that the fidelity quantum kernel is a projection of a $U$-transformed and $U$-inverse transformed ($U^{\dagger}=U^{-1}$ for unitary operators) state, $U^{\dagger}(y) U(x) \rho_0 U^{\dagger}(x)U(y)$, on the initial state $\rho_0$. The schematic representation of this process is shown in Fig.~\ref{fig:FQK_idea}. Therefore, the circuit for a fidelity quantum kernel estimation in the Loschmidt echo approach consists of: 
\begin{enumerate}
    \item[i] preparation of the initial state $\rho_0$;
    \item[ii] application of the first data point embedding, $U(x)$;
    \item[iii] application of the inverse embedding of the second data point $U^{\dagger}(y)$; 
    \item[iv] estimation of the probability of obtaining the initial state after the transformations, $P(| 0 \rangle) = | \langle \phi(y) | \phi(x) \rangle |^2 = K_{FQ}(x,y)$.
\end{enumerate}

In quantum computing, obtaining the final quantum state is not the end of the computation, as we still need to extract classical, human-readable information from the quantum system. This is done in a measurement process. During measurement, the state of the quantum system collapses into one of the states of the measurement basis with the probability according to the Born rule~\cite{born1926quantenmechanik}.
The value of the fidelity quantum kernel is effectively the probability of the state $|0\rangle$ in a superposition state $U^{\dagger}(y) U(x) \rho_0 U^{\dagger}(x)U(y)$. Thus, measuring the final state of the Loschmidt echo circuit (Fig.~\ref{fig:FQK_idea}b.) multiple times allows us to approximate the value of the fidelity quantum kernel. This process is called a quantum kernel estimation, and in the case of the fidelity quantum kernel is fully equivalent to the binomial proportion estimation for the distribution $\text{Bin}(k, S; p=K_{FQ}(x,y))$, where $k$ is the number of times we measured the initial state and $S$ is the number of circuit run repetitions.

\subsection{Kernel Bandwidth}
The exponentially large target feature space induced by quantum kernels may negatively affect the performance of subsequent kernel methods.
\ac{QML} is prone to the effects of the \textit{curse of dimensionality} \cite{mccleanBarrenPlateausQuantum2018}.
It manifests itself in quantum kernel methods as an exponential kernel concentration~\cite{thanasilpExponentialConcentrationQuantum2024}, which affects the trainability of the \ac{ML} model. Clarifying, as the number of qubits in our model increases, the kernel values tend to concentrate around a constant value exponentially quickly. The mean kernel value tends to a constant value, whereas the variance vanishes. That, together with the inherent granularity of the measurement process, leads to indistinguishability of different kernel values unless we perform an exponential number of circuit runs.

In fidelity quantum kernels, the values tend to statistically vanish---a simple explanation for this effect exists. As we increase the number of qubits, the dimension of the available Hilbert space increases to $2^n$. Then, if we randomly sample two quantum states in such a Hilbert space with many orthogonal directions, the mean fidelity between the states scales as $2^{-n}$, and similarly the standard deviation scales as $2^{-n}\sqrt{(1-2^n)/(1+2^{-n})}$. Thus, one can end up with a trivial model in which the predictions on samples are independent of the input data. 
Furthermore, the high dimension of the Hilbert space reduces the inductive bias of the models~\cite{kubler2021inductive}. It has been found that to learn from data, one cannot successfully use the full expressive power of the quantum Hilbert space, but rather restrict it. Otherwise, one has to provide an exponentially large training set.
Therefore, both of the above effects can be mitigated by constraining the target Hilbert space.
One proposition is to perform quantum computation on the whole Hilbert space, but limit the measurement process to only a subset of qubits. That gives rise to the variety of projected quantum kernels~\cite{huangPowerDataQuantum2021}. 
However, in order to simulate the projected kernels, one needs to obtain a whole system's density matrix, which is out of the question for large quantum kernels.
Another approach is to introduce a bandwidth, a procedure known from classical kernel computation methods~\cite{kohler2014review}.
Effectively, this approach consists of data rescaling by a small parameter $c$, $x \mapsto c\cdot x$. 
In the classical RBF kernel function, defined as
\begin{equation}
    K_{RBF}(x,y) = e^{-\gamma||x-y||^2},
\end{equation}
one can recognize the bandwidth parameter as $c=\sqrt{\gamma}$. Changing its value determines the spread of the above Gaussian. For large $\sqrt{\gamma}$, the kernel is sharply peaked and might overfit the data, while, on the contrary, small $\sqrt{\gamma}$ can lead to underfitting. 
A similar idea can be applied to quantum kernels. Assuming that the range of data values $x$ covers all possible distinct unitary transformations generated in the circuit, we expect that, with a large number of qubits, our models will suffer from the problems mentioned above, leading to extreme overfitting.
Assuming that the quantum states generated by $U(x)$ cover the full image of the transformation, introducing the bandwidth parameter $U(c\cdot x)$ with $c<1$ restricts the set of states accessible through the transformation, thereby limiting the expressivity of the model. In fact, the described behavior has been reported in the literature~\cite{shaydulin2022importance, canatar2022bandwidth}.

\begin{figure*}[ht]
    \begin{subfigure}[b]{\linewidth}
        \centering
        \includegraphics[width=\linewidth]{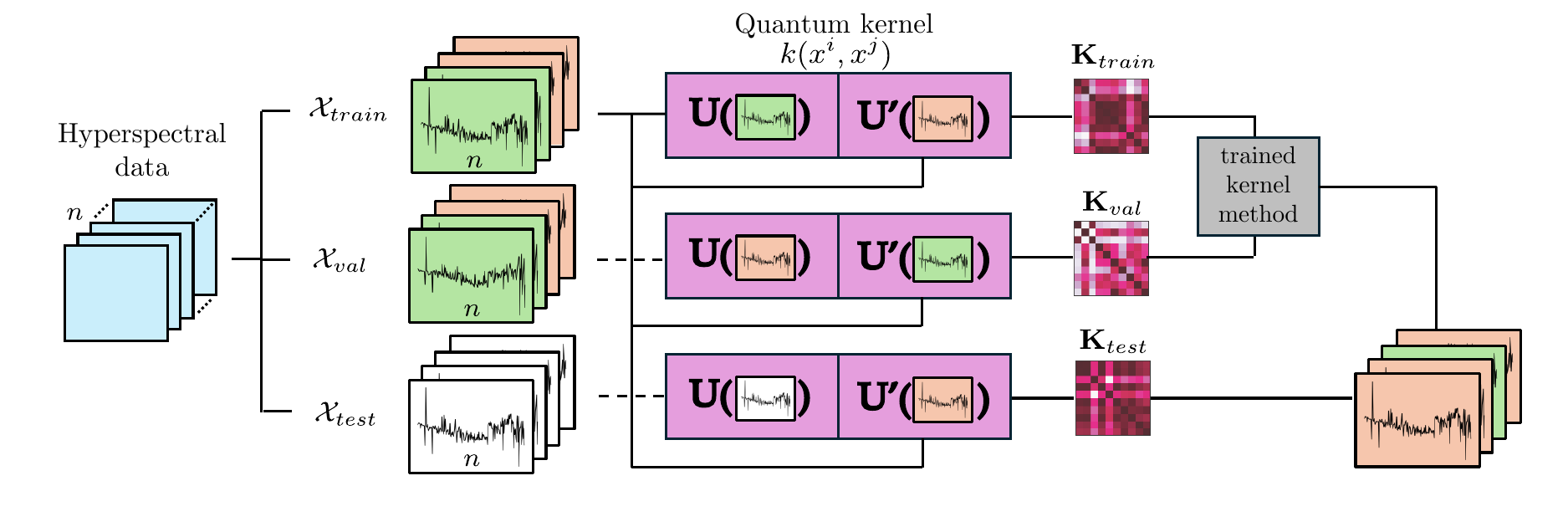}
        \caption{}
    \end{subfigure}
    \vfill
    \begin{subfigure}[b]{\linewidth}
        \centering
        \[
        \begin{array}{c}
            \Qcircuit @C=0.55em @R=1em {
            \lstick{\ket{0}} & \gate{H} & \gate{R_{z}(x^i_1)} & \gate{R_{y}(x^i_1)} &
            \ctrl{1} & \qw & \qw & \qw & \gate{R_{z}(x^i_1)}
            \ar@{.}[]+<2.1em,1.5em>;[d]+<2.1em,-8em> &
            \gate{R_{z}(x^j_1)} & \qw & \qw & \qw & \ctrl{1} &
            \gate{R_{y}(x^j_1)} & \gate{R_{z}(x^j_1)} & \gate{H} & \meter \\
            \lstick{\ket{0}} & \gate{H} & \gate{R_{z}(x^i_2)} & \gate{R_{y}(x^i_2)} &
            \targ & \ctrl{1} & \qw & \qw & \gate{R_{z}(x^i_2)} &
            \gate{R_{z}(x^j_2)} & \qw & \qw & \ctrl{1} & \targ &
            \gate{R_{y}(x^j_2)} & \gate{R_{z}(x^j_2)} & \gate{H} & \meter \\
            \lstick{\ket{0}} & \gate{H} & \gate{R_{z}(x^i_3)} & \gate{R_{y}(x^i_3)} &
            \qw & \targ & \ctrl{1} & \qw & \gate{R_{z}(x^i_3)} &
            \gate{R_{z}(x^j_3)} & \qw & \ctrl{1} & \targ & \qw &
            \gate{R_{y}(x^j_3)} & \gate{R_{z}(x^j_3)} & \gate{H} & \meter \\
            \lstick{\vdots} & & & & & & & & & & & & & & &      \\
            \lstick{\ket{0}} & \gate{H} & \gate{R_{z}(x^i_d)} & \gate{R_{y}(x^i_d)} &
            \qw & \qw & \qw & \targ \qwx & \gate{R_{z}(x^i_d)} &
            \gate{R_{z}(x^j_d)} & \targ \qwx & \qw & \qw & \qw &
            \gate{R_{y}(x^j_d)} & \gate{R_{z}(x^j_d)} & \gate{H} & \meter \\
            }
    \end{array}
    \]
        \caption{}
    \end{subfigure}

    \caption{(a) Overall scheme of the implemented method, used to evaluate quantum kernels. The classification of unseen test spectra $\mathcal{X}_{test}$ is performed by training a kernel method on $\mathcal{X}_{train}$ and optimizing its parameters to maximize accuracy on $\mathcal{X}_{val}$. (b) Quantum circuit used for kernel computation in our method. It follows the fidelity quantum kernel approach shown in Fig. \ref{fig:FQK_idea}, the vertical dotted line marking the separation between $U(x^i)$ and $U^{-1}(x^j)$. Each feature $x^i_n$ is used as a parameter for three rotation gates, i.e., $R_z, R_y, R_z$, applied to each qubit $n$, for a total of $d$ qubits. $H$ represents the Hadamard gate, and $\oplus$ represents the CNOT gate, here applied to each qubit $n\geq2$ and controlled by the state of the qubit $n-1$.}
    \label{fig:QK_circuit}
\end{figure*}

\section{Large-scale Quantum Kernels for Hyperspectral Data Classification}

In this section, we present a procedure for evaluating the performance of quantum kernels in classifying data generated by hyperspectral sensors. First, we introduce a mathematical formulation for pixel-wise classification of hyperspectral data. Then, we propose a way to define and simulate large-scale quantum kernels.

\subsection{Pixel-wise Classification}

Let $\mathcal{X} = \{ x_i \in \mathbb{R}^d \mid i = 1, \ldots, n \}$ represent the set of hyperspectral pixel vectors, where each $x_i$ is a feature vector with $d$ spectral bands, corresponding to the pixel $i$. In a supervised setting, each pixel has an associated label $y_i \in \mathcal{Y}$ among $C$ possible labels, where $C$ is the number of target classes. For our exploratory analysis, we train an SVM on two-class problems to provide a more focused examination of the generated kernels. The method is then extended to a multiclass setting by employing a 1-vs-rest classification strategy.

From $\mathcal{X}$, we generate the splits $\mathcal{X}_{train}$, $\mathcal{X}_{val}$, $\mathcal{X}_{test}$ of size $N_{train}, N_{val}, N_{test}$, respectively. As the names suggest, $\mathcal{X}_{train}$ is used for the SVM training, $\mathcal{X}_{val}$ for validation (i.e., tuning the model's hyperparameters), and $\mathcal{X}_{test}$ for testing and quantifying its generalization abilities.

We can compute the train, validation, test kernel matrices $\mathbf{K}_s \in \mathbb{R}^{N_{train} \times N_{s}}$, $s=\{train,val,test\}$, with entries $K_{ij} = K(x_i, x_j)$. A pixel-wise classifier $f: \mathbb{R}^d \rightarrow \mathcal{Y}$ is defined as:
\begin{equation}
f(x) = \mathrm{sign}\left( \sum_{i=1}^{N_{train}} \alpha_i y_i K(x_i, x) + b \right),
\end{equation}
where $\alpha_i$ and $b$ are the model's parameters learned during training. For unseen test pixels $\{ x_j^* \}_{j=1}^{m}$, the predicted labels are given by:
\begin{equation}
\hat{y}_j = f(x_j^*) \qquad \text{for } j = 1, \ldots, m.
\end{equation}

\subsection{Quantum Kernel Simulation}
\label{sub:qks}


Our goal is to use quantum kernels to implement the kernel function $K(x_i, x_j)$ and compute the kernel matrices $\mathbf{K}_s$. Figure~\ref{fig:QK_circuit} presents an overview of the proposed pipeline together with the adopted quantum circuit ansatz. The feature map employs a rotation-based encoding, where each spectral band is mapped to a dedicated qubit via parameterized single-qubit rotations. This one-to-one correspondence preserves the full spectral resolution of hyperspectral pixels and avoids any dimensionality reduction or feature aggregation prior to quantum processing, which is essential for analyzing quantum kernels in high-dimensional regimes. A shallow, linear entanglement pattern implemented through CNOT gates introduces correlations between neighboring spectral channels, reflecting the smooth and locally correlated nature of hyperspectral signatures while maintaining a controlled level of expressivity. Importantly, the ansatz is fixed and does not involve additional trainable quantum parameters or adaptive circuit structure. This design choice ensures that learning is governed entirely by the induced kernel, its bandwidth, and the classical SVM regularization, enabling a stable and interpretable kernel-level analysis without navigating the highly nonconvex training landscapes typically associated with variational quantum models. Notably, the number of qubits equals the number of spectral features, $n$, which is typically on the order of hundreds for hyperspectral data. While this precludes execution on current quantum hardware and renders standard statevector simulation infeasible due to exponential memory scaling, the tensor-network-based simulation strategy adopted in this work allows us to investigate this class of quantum kernels at scales relevant to real-world hyperspectral datasets~\cite{willsch2019benchmarking}.

In general, \texttt{statevector} simulation for fidelity quantum kernel estimation scales exponentially with the number of qubits, $n$, $\mathcal{O}(2^n)$. Moreover, computing the entire kernel matrix, which contains pairwise similarities between data points, scales quadratically with the dataset size ($\mathcal{O}(N^2)$). Thus, quantum fidelity kernels are hard to simulate classically.
However, the recent work~\cite{chen2025validating}, building on the \texttt{cuQuantum SDK}~\cite{bayraktar2023cuquantum}, challenges this view. The authors note that for fidelity quantum kernel estimation, only the probability of obtaining the vacuum state $|0\rangle$ is required, rather than the whole state vector. It uses a framework based on tensor network contraction, which has been proven to be able to efficiently simulate $T$-gate quantum circuits with low entanglement~\cite{markov2008simulating}.
Leveraging optimized contraction path identification, path reuse, and parallelization via Message Passing Interface (MPI), they achieve a quadratic scaling $\mathcal{O}(n^2)$ in the number of qubits and effectively constant scaling $\mathcal{O}(1)$ in the dataset size, without any loss of simulation accuracy. Their results are demonstrated for block-encoded states with a linear entangling strategy.

In this work, we adopt this approach, verify the reported performance, and explore its potential for high-qubit \ac{QML}, where the number of qubits matches the feature dimensionality of remote sensing datasets.


\section{Experimental Setup}
\label{sec:exp}

In this section, we describe the experimental setup considered for evaluating high-dimensional quantum kernels. We explain how we created the training splits, how we compare the results between quantum and classical methods, as well as how we implemented the methods in practice.

\subsection{Datasets}
\label{sub:datasets}

The experimental validation of the quantum kernel methodology is performed on supervised learning tasks based on hyperspectral data. In particular, we create balanced and normalized training-validation-test splits taken from two hyperspectral datasets. To study the performance of our model as the number of features $n$ increases, we rank them according to their variance and select the first $n$ features (with the largest variance).
No feature extraction is performed, since the goal is to work with the original data (in terms of the number and characteristics of spectral bands) as closely as possible, ultimately considering the whole spectral signature.
In our work, we use quantum embedding maps that employ the same number of qubits as features, so there is no conflict in notation for the symbol $n$. 

\subsubsection{Indian Pines}
Indian Pines \cite{PURR1947} is a widely used hyperspectral benchmark acquired by the Airborne Visible/Infrared Imaging Spectrometer (AVIRIS) sensor over agricultural fields in Northwestern Indiana, USA. It consists of a labeled $145\times145$ pixel scene with $220$ spectral bands spanning the $0.4–2.5 \mu m$ wavelength range. After removing bands affected by water absorption, $200$ bands are typically retained for analysis.
The work of Nalepa et al.~\cite{nalepa2019validating} provides guidelines for creating training splits from the Indian Pines dataset. 
We generate a training split for each of the four splits provided in the aforementioned paper, ensuring spatial separation between training and test data. We define two learning tasks:

\begin{itemize}
    \item \textit{Binary classification} task. For each split, we randomly sample 50 training, 50 validation, and 100 test instances taken from 2 classes, i.e., \textit{corn-mintill} (class $3$), \textit{soybean-notill} (class $10$). This serves as the main test bed for our experiments, in which we perform an extensive kernel analysis.
    \item \textit{Multiclass classification} task. For each split, we randomly sample 100 training, 100 validation, and 200 test instances taken from 4 classes, i.e., \textit{corn-mintill} (class $3$), \textit{soybean-notill} (class $10$), \textit{hay-windrowed} (class $8$), and \textit{grass-pasture} (class $5$).
\end{itemize}

To ensure reproducibility, the data used in this work are provided in the public repository~\cite{ourcode2025}.
The Indian Pines classes considered in the binary and multiclass experiments (corn‑mintill, soybean‑notill, hay‑windrowed, and grass‑pasture) were selected based on two criteria. First, they contain a sufficient number of samples to allow stable subsampling and statistically meaningful evaluation, avoiding classes whose very small size leads to high variance and split‑dependent results. 
Second, we choose these classes because the CNN-based methods included in \cite{nalepa2019validating} show a higher variability in cross-validated per-class segmentation accuracies, hinting to a non-triviality for the classification task.
We note that the multiclass setting is restricted to four categories in order to keep the computational cost of quantum‑kernel matrix construction tractable across all splits, since extending the analysis to all well‑represented Indian Pines classes would require substantially larger kernel evaluations with quadratic scaling in the number of samples per one‑vs‑rest classifier.

\subsubsection{Methane Detection}

Methane Dataset~\cite{Thompson2017} is a hyperspectral dataset designed for methane source detection and segmentation. It was acquired using the Airborne Visible-Infrared Imaging Spectrometer-Next Generation (AVIRIS-NG) sensor over the Four Corners Methane Hot Spot in New Mexico, USA, during April 19–21, 2015. The dataset contains hyperspectral images with over 400 spectral bands in the 400-2500\,nm range and a spatial resolution of 2.8\,m. 
The dataset comprises 178 patches (512$\times$512 pixels each) extracted from 27 flights. Each flight is accompanied by a CH$_4$ enhancement map (mag1c map) generated using the mag1c tool~\cite{2020_Foote}, which serves as the basis for manually annotated binary ground truth, distinguishing methane super-emitters and their plumes from the background. Each patch includes a header file (with date and geographic coordinates), hyperspectral image, CH$_4$ enhancement map, and binary ground truth. Among these, 89 patches contain methane emitters. From each patch, a representative pixel was extracted to construct spectral curves. For patches containing methane, the pixel with the smallest absolute difference from the median methane value in the mag1c map was selected. For background patches, the pixel closest to the median value of the mag1c map across the entire image was chosen. This procedure yielded 178 spectral curves, i.e., 89 representing methane and 89 representing no-methane areas. Each curve consists of $428$ spectral bands and was labeled accordingly: those extracted from methane patches were labeled as \textit{methane}, while those from no methane patches were labeled as \textit{background}. We consider all available $428$ spectral bands and a total of $5$ splits using $5$-fold cross-validation on a balanced set of $178$ spectral curves. The data is partitioned using a fixed $2{:}1{:}2$ ratio for training, validation, and test sets, resulting in $71$ training, $36$ validation, and $71$ test instances per fold. This setup ensures that each fold contains a representative and balanced mixture of both classes, facilitating robust evaluation and fair comparison of model performance across different splits.


\subsection{Model Training}
\label{sub:model_training}

We stress that a meaningful comparison between classical and \ac{QML} models can be challenging. In the \ac{QML} literature, two approaches to such a comparison exist; either it is built on learning performance metrics (e.g., accuracy, F1 score) between the quantum model and the state-of-the-art classical method~\cite{9883992, sebastianelli2021circuit, sebastianelli2023quantum, miroszewski2023detecting}, or more commonly between the quantum models and its \textit{classical counterpart}~\cite{wijata2024detection, guptaPotentialQuantumMachine, delilbasic2021quantum, delilbasic2023single, otgonbaatar2021assembly}.
The former overlooks the key differences between quantum and classical models, instead focusing on their shared goal. Although the progressive maturity of the field of \ac{QML} will increase the relevance of this approach, we also highlight the current necessity for the latter approach. Being able to compare analogous models belonging to classical and quantum \ac{ML} paves the way for understanding the crucial differences between the fields. Exploiting such differences is not only an ultimate goal in obtaining a quantum advantage, but also a prospect for extending the limits of useful computation in general.

In this work, we avoid defining how to find classical counterparts to general QML models. We point out that in the case of kernel methods, such an analogy can be provided straightforwardly. As mentioned in Section~\ref{sec:QKMs}, quantum kernel methods extend the set of efficiently computable similarity measures which can then be used in classical \ac{ML} routines.
Therefore, when comparing quantum and classical kernel methods, one should focus on a faithful analysis of accessible kernel functions. In general, the analysis should take into account the cost of obtaining classical and quantum kernels~\cite{miroszewski2024searchquantumadvantageestimating}. However, in the case of this work, based on recent advances in classical simulation of quantum kernels~\cite{chen2025validating}, we rather focus on providing the models with an equivalent number of hyperparameters. This issue has been customarily overlooked in the literature, but recently is gaining more and more attention (see~\cite{mauro2025quantum}).

In summary, we compare an RBF kernel and a quantum kernel, each of which includes two hyperparameters. For the RBF kernel, these are the spread of a Gaussian ($\gamma$), effectively controlling the kernel bandwidth, and the penalty term in the SVM routine ($C$). For the quantum kernel, those are the bandwidth parameter ($c$) and a penalty term hyperparameter, $C$. Hence, the comparison is meaningful in the sense described above. We find the optimal hyperparameters to train our model on $\mathcal{X}_{train}$ by running Bayesian optimization~\cite{snoek2012practical} with 50 iterations, where the objective function is the accuracy of our model on $\mathcal{X}_{val}$. Additionally, in order to illustrate the issues connected with high-dimensional, multi-qubit systems, we report on the quantum model without the bandwidth hyperparameter optimization ($c=1$). The classical versus quantum kernel comparison for binary classification follows the latter approach and serves as a basis for an in-depth analysis of the behavior of quantum kernels in high-dimensional Hilbert spaces.

In addition, we perform a former (metric-based) analysis to compare quantum kernels with state-of-the-art methods for pixel-wise hyperspectral data classification  in the small-sample regime.
The comparison is presented with respect to multiple performance metrics. We further extend this analysis to multiclass classification on the Indian Pines dataset, adopting a one-versus-rest strategy within the SVM framework.

Deep learning models, including convolutional neural networks and Transformer-based architectures, currently represent the dominant paradigm in hyperspectral image classification, particularly in spatial–spectral settings with large labeled datasets \cite{chen2016deep, hong2021spectralformer, paoletti2019deep}. These approaches typically operate on image patches rather than individual pixels and rely on extensive parameter learning and strong spatial priors to achieve state-of-the-art performance. In contrast, the present work focuses on a pixel-wise, spectral-only classification regime with limited labeled samples, where model capacity, inductive bias, and generalization properties can be analyzed in a controlled manner. In such a setting, direct experimental comparison with deep neural networks would be inherently unfair and potentially misleading, as the learning objectives, data representations, and sample efficiency requirements differ substantially. We therefore position quantum kernel methods as complementary approaches, particularly suited for high-dimensional, small-sample scenarios, and restrict quantitative baselines to kernel-based and lightweight classical models. For completeness, we note that CNN- and Transformer-based methods have demonstrated strong performance in large-sample spatial–spectral regimes, as reported in numerous works in the hyperspectral imaging literature; however, these results reflect a different problem formulation than the one considered here. 

\subsection{Implementation}

The quantum kernel circuit in Fig.~\ref{fig:QK_circuit} is implemented using the \texttt{Qiskit} library and then converted to a tensor network formulation with \texttt{cuTensorNet}. To evaluate the advantage of this approach, we compare its runtime with the \texttt{statevector} quantum simulator in \texttt{Qiskit}, without circuit conversion. The experiments were performed on 12 nodes from the DEEP cluster located at Forschungszentrum Jülich~\cite{eicker2011deep}. Each node is equipped with 2 Intel Xeon Platinum 8260M CPUs and an NVIDIA V100 GPU. Data parallelism is employed for kernel matrix computation, with inter-node communication being handled with MPI. Such a setup is necessary to conduct the full-scale experiments reported in this paper, which collectively require around a week. Individual runs can be performed with fewer resources or without HPC, provided that a GPU is available.

To ensure reproducibility, the public repository~\cite{ourcode2025} includes the open-source implementation used for our simulations, along with the data, results, and analysis code.

\section{Results}
\label{sec:results}
The delineated experimental setup provides a variety of results for each dataset and for each training split. In this section, we focus on the following aspects.

\begin{figure}[t]
    \centering
    \includegraphics[width=0.8\linewidth]{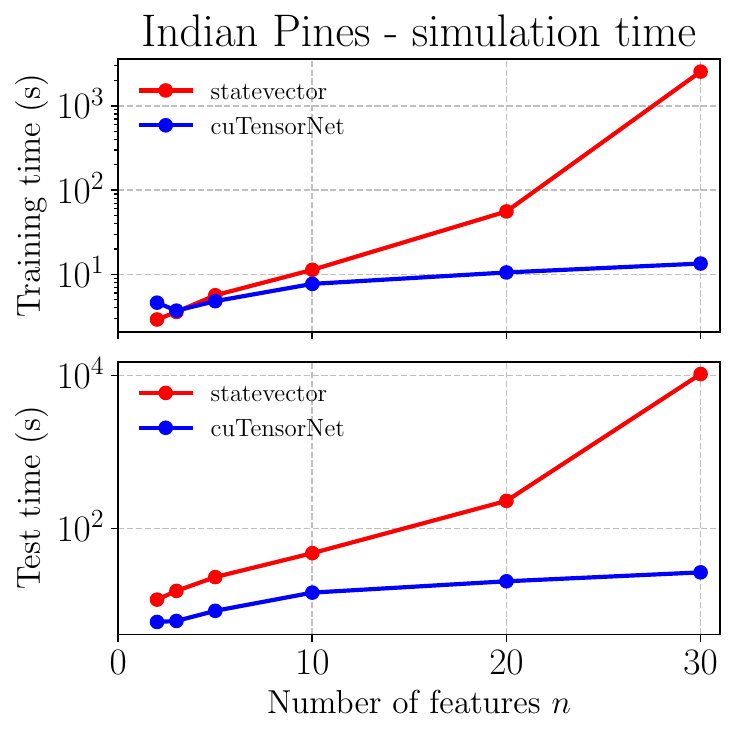}
    \caption{Indian Pines---run time for \texttt{statevector} and \texttt{cuTensorNet} quantum circuit simulation (up to $30$ qubits) for generating $\mathbf{K}_{train}$ and $\mathbf{K}_{test}$ for a single Indian Pines training split. We highlight that the computational workload is the same for each split, thus it is representative.}
    \label{fig:time_indian}
\end{figure}

\begin{figure}[t]
    \centering
    \includegraphics[width=0.9\linewidth]{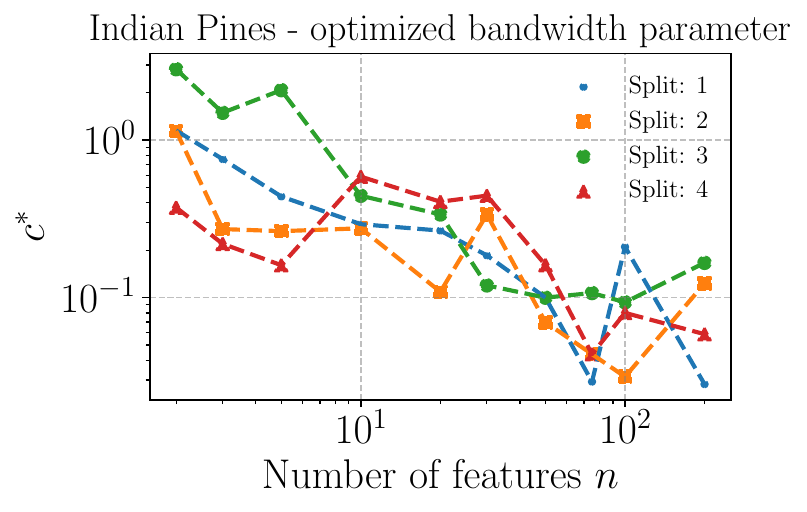}
    \caption{Indian Pines---optimized bandwidth parameter $c^*$ found in the validation stage during the experiment described in the caption of Fig.~\ref{fig:ACC_indian} as a function of the number of features $n$. The optimization strategy is described in Section~\ref{sub:model_training}. Table~\ref{tab:fits_indian} gathers the power-law and exponential fits of the above data. 
    }
    \label{fig:c_optimal_indian}
\end{figure}

\begin{figure*}[!ht]
    \centering
    \includegraphics[width=\textwidth]{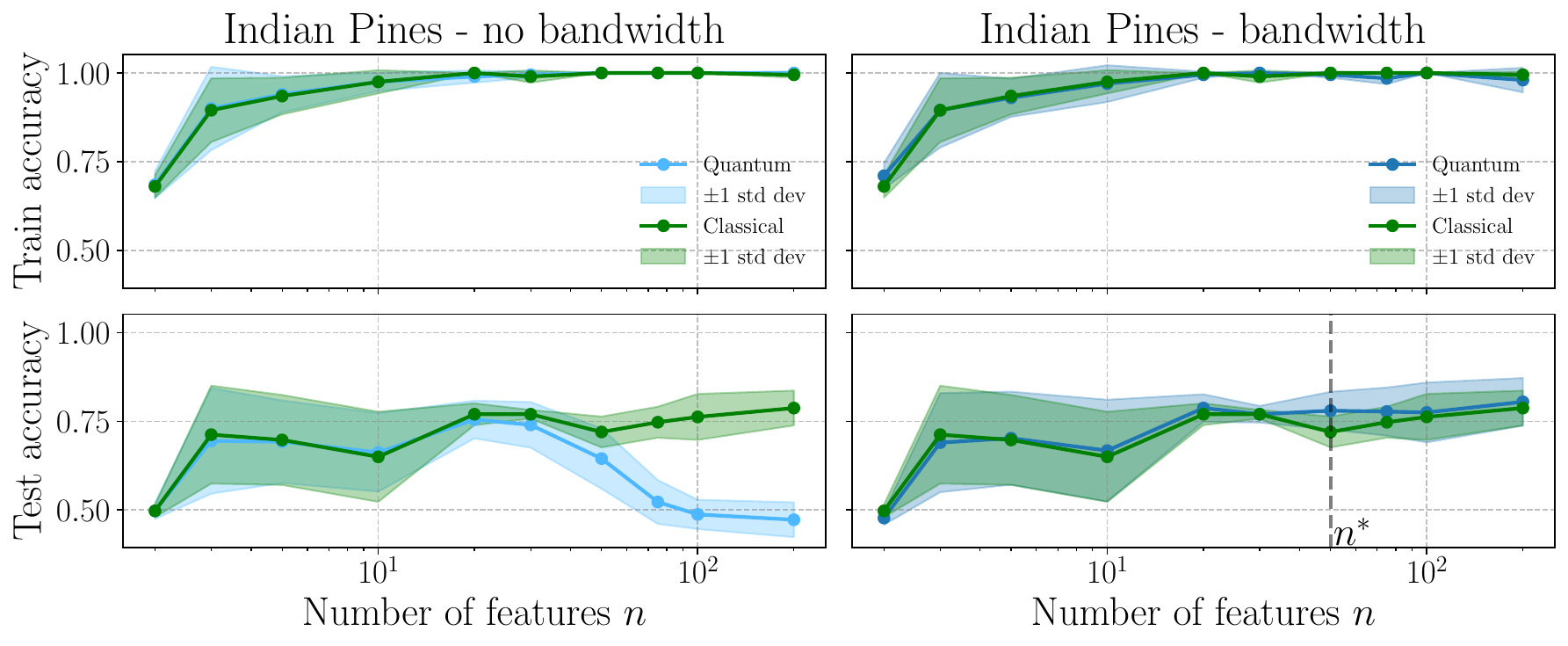}
    \caption{Indian Pines---average train (\textit{top}) and test (\textit{bottom}) accuracy for binary classification of the \textit{corn-mintill} (class $3$) and \textit{soybean-notill} (class $10$) data points from the Indian Pines dataset as a function of the number of features $n$ used. The legend distinguishes between the \textit{quantum} model (described in Section~\ref{sub:qks}) and the \textit{classical} model (described in Section~\ref{sub:model_training}).}
    \label{fig:ACC_indian}
\end{figure*}

\begin{figure*}[!ht]
    \centering
    \includegraphics[width=\textwidth]{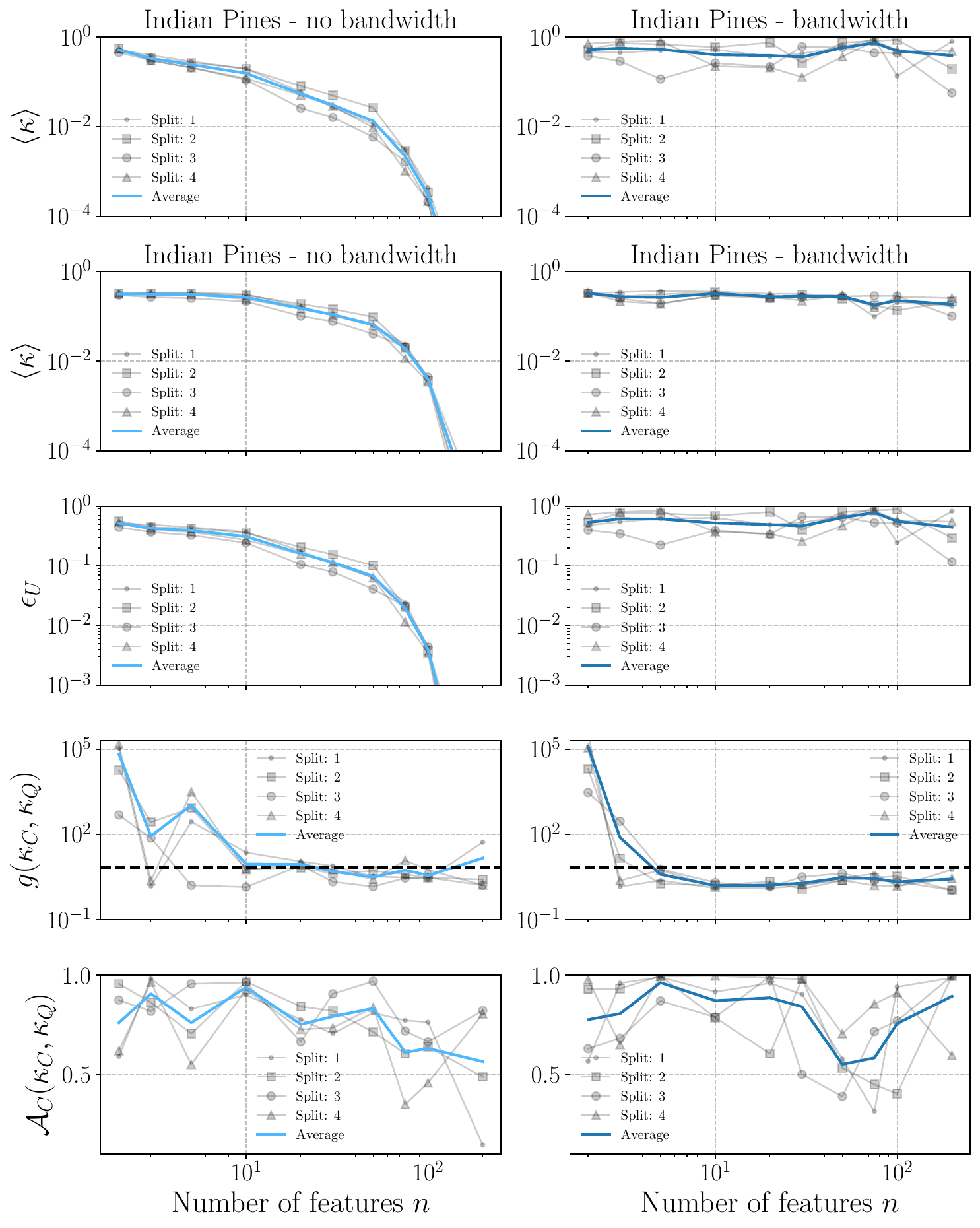}
    \caption{Indian Pines---aggregate plots for the analysis of kernel matrices: mean kernel values $\langle \kappa \rangle$, std of kernel values $\sigma( \kappa)$, expressibility $\epsilon$, geometric difference between classical and quantum matrices $g$, alignment between classical and quantum matrices $\mathcal{A}$.}
    \label{fig:kernel_analysis_indian}
\end{figure*}

\begin{figure*}[!ht]
    \centering
    \includegraphics[width=0.9\textwidth]{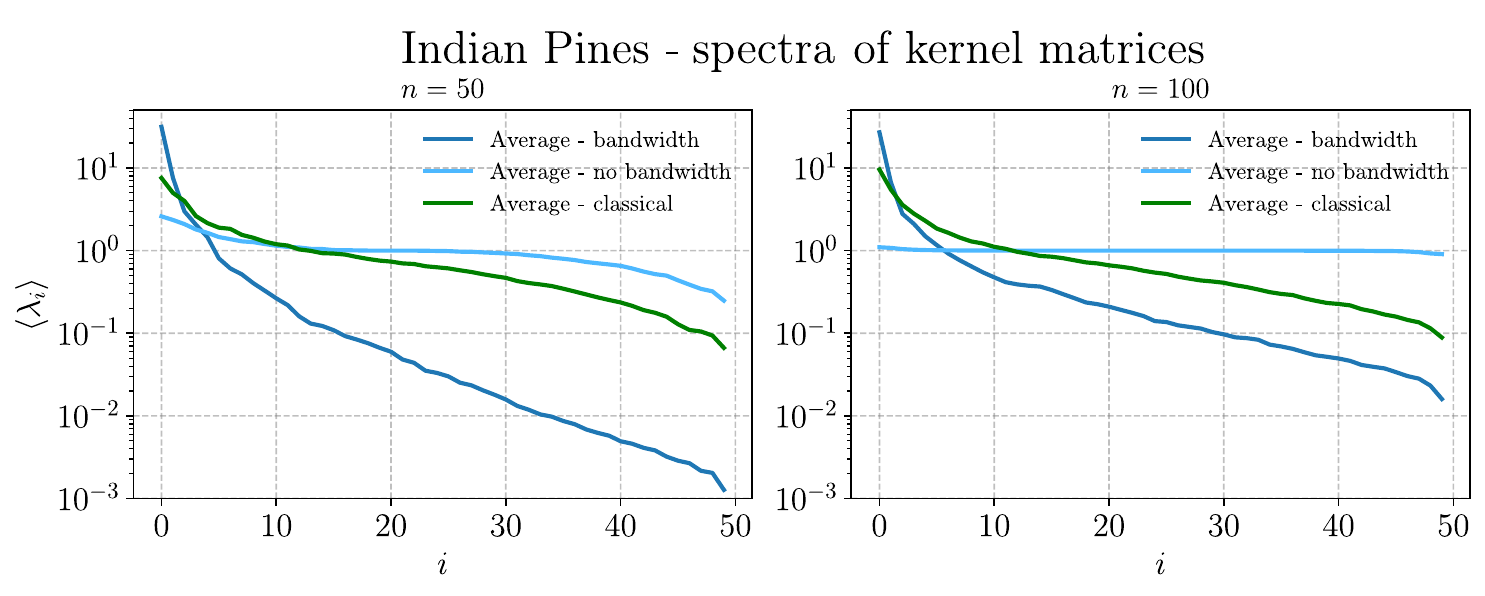}
    \caption{Indian Pines---spectra of the training kernel matrices $\mathbf{K}_{train}$, averaged on all 4 training splits, for selected number of features $n=\{50,100\}$}
    \label{fig:indian_spectra}
\end{figure*}

\begin{enumerate}
    \item We show a runtime comparison between the calculation of quantum kernel matrices $\mathbf{K}_{train}, \mathbf{K}_{test}$ using \texttt{statevector} and \texttt{cuTensorNet} quantum circuit simulation (Fig.~\ref{fig:time_indian} and Fig.~\ref{fig:time_methane}), estimating the computational complexity of the two approaches and highlighting the advantage of the tensor network contraction.
    \item We show the optimized values of the quantum kernel bandwidth parameter $c$ and extrapolate suitable fits for them, comparing them with expected scaling laws (Fig.~\ref{fig:c_optimal_indian} and Fig.~\ref{fig:c_optimal_methane}; Table~\ref{tab:fits_indian} and Table~\ref{tab:fits_methane}).
    \item We compare training and test accuracy of an SVM trained with classical and quantum kernels, with (\textit{bandwidth}) and without (\textit{no bandwidth}) optimizing $c$ (Fig.~\ref{fig:ACC_indian} and Fig~\ref{fig:ACC_methane}). This is a sufficient performance metric, as the selected splits are class-balanced.
    \item We present performance comparisons of quantum kernel-based SVMs with state-of-the-art methods for binary classification (Table~\ref{tab:Indian_comparison_results} and Table~\ref{tab:Methane_comparison_results}) and multiclass classification (Table~\ref{tab:Indian_multiclass_comparison_results}).
    \item We include additional analysis of the obtained training kernel matrices $\mathbf{K}_{train}$, aligning our results to quantum kernel theory . We show the mean $\langle \kappa \rangle$ and the standard deviation $\sigma( \kappa)$ of kernel values, 
    expressibility $\epsilon$, geometric difference $g$, alignment $\mathcal{A}$ between classical and quantum kernel matrices, as well as their spectra (Fig.~\ref{fig:kernel_analysis_indian}, Fig.~\ref{fig:indian_spectra}, Fig.~\ref{fig:kernel_analysis_methane} and Fig.~\ref{fig:spectra_methane}).
\end{enumerate}

In the following, we describe each of these metrics and outline why they are important for our analysis.
Plotting the mean kernel values $\langle \kappa \rangle$ provides a statistical characterization of the kernels as a function of the number of qubits $n$.
In~\cite{thanasilpExponentialConcentrationQuantum2024}, the authors predict that, when using the Loschmidt echo method for fidelity quantum kernel estimation, the individual kernel values will statistically concentrate towards zero as the number of qubits increases.
Therefore, plotting the mean kernel value and, similarly, the standard deviation allows us to identify whether the exponential concentration problem occurs.
Calculating expressibility $\epsilon$ in a way described in~\cite{holmes2022connecting,nakaji2021expressibility,sim2019expressibility} enables the assessment of the expressibility measures, i.e., how extensively a quantum feature map can represent data in the Hilbert space. 
More formally, expressibility quantifies the ability of the feature map $U(x)$ to generate a diverse set of quantum states $|\psi(x)\rangle$ across the Hilbert space as the input data $x$ varies.
It is assessed through the distribution of pairwise state fidelities, with higher expressibility (low $\epsilon$ value) corresponding to states that approximate the Haar-random distribution, and lower expressibility (high $\epsilon$ value) indicating a more restricted subset.
Additionally, we analyze the two measures of the mutual relations between classical and quantum kernels.
The geometric difference $g(\kappa_C \| \kappa_Q)$, introduced in~\cite{huangPowerDataQuantum2021}, is a quantitative way to compare two kernel matrices, such as one from a classical \ac{ML} model $\kappa_C$ and one from a quantum model $\kappa_Q$. It has been found that for $g(\kappa_C \| \kappa_Q) \propto \sqrt{N}$, there exists a potential for the prediction advantage in a quantum model. In other cases, the classical model is likely to be competitive with the quantum model, or the learning task is hard for both.
Kernel alignment $\mathcal{A}(\kappa_C, \kappa_Q)$ is a normalized Frobenius product between kernel matrices. Interpreting kernel matrices as vectors, $\mathcal{A}$ can be seen as a cosine of the angle between those vectors. Thus, it can be used as a measure of the similarity of matrices.
The inductive bias of a kernel method refers to its inherent preference for specific classes of functions, determined by the geometry of the feature space implicitly induced by the kernel. A standard way to characterize this bias is through the spectral analysis of the kernel matrix, whose eigenvalues quantify the effective dimensionality and complexity of the induced representation. As established in classical kernel theory, the decay of the Gram matrix spectrum governs how strongly a kernel favors smooth, low-complexity functions versus more expressive hypotheses, with rapidly decaying eigenvalues indicating a strong bias toward simpler patterns and slower decay corresponding to a higher effective capacity of the model \cite{scholkopf2002learning, shawe2004kernel}. This spectral perspective has been shown to play a central role in understanding generalization behavior and inductive bias in kernel methods, including both classical and quantum settings, where differences in spectral decay directly reflect differences in the structure of the induced feature spaces \cite{kubler2021inductive, canatar2021spectral}. In the following, we therefore use the spectra of the kernel matrices as a tool to interpret and compare the behavior of classical and quantum kernels within the established framework of kernel learning theory.

\begin{table}[t]
\caption{Indian Pines---fit parameters and $R^2$ value of the power law ($log(c^*) = a \cdot log(n) + b$) and exponential ($log(c^*) = a \cdot n + b$) fits to the optimal bandwidth parameters $c^*(n)$. The found bandwidth parameters $c^*$ are presented in Fig.~\ref{fig:c_optimal_indian}.
}
\label{tab:fits_indian}
\centering
\begin{tabular}{clrrr}
\toprule
 Split &         Fit &       a &       b &    $R^2$ \\
\midrule
     1 &   Power law & $-0.6558$ &  $0.4183$ & $0.7348$ \\
     1 & Exponential & $-0.0220$ & $-0.6784$ & $0.5042$ \\
     \arrayrulecolor{gray!50}\midrule
     2 &   Power law & $-0.7025$ &  $0.1310$ & $0.7662$ \\
     2 & Exponential & $-0.0281$ & $-0.8930$ & $0.7513$ \\
     \arrayrulecolor{gray!50}\midrule
     3 &   Power law & $-0.9462$ &  $1.6487$ & $0.9339$ \\
     3 & Exponential & $-0.0323$ &  $0.0852$ & $0.6643$ \\
     \arrayrulecolor{gray!50}\midrule
     4 &   Power law & $-0.3287$ & $-0.6505$ & $0.2953$ \\
     4 & Exponential & $-0.0183$ & $-0.9607$ & $0.5606$ \\
\arrayrulecolor{black}\bottomrule
\end{tabular}
\end{table}

\begin{table*}[]
    \centering
    \caption{Indian Pines---binary classification performance metrics of small-sample state-of-the-art baseline ML models and the presented quantum kernel SVM. The models are trained on 4 splits containing 50-50-100 train-validation-test data points from classes \textit{corn-mintill} (class $3$) and \textit{soybean-notill} (class $10$). Results are for number of features $n=50$, reported as mean ± standard deviation over 4 splits. Best results for each metric are highlighted in bold.}
    \label{tab:Indian_comparison_results}
\begin{tabular}{lccccc}
\toprule
Model &           Accuracy &          Precision &             Recall &        Specificity &          $\text{F}_\text{1}$ \\
\midrule
AdaBoost \cite{freund1997decision}               &  0.672 $\pm$ 0.084 &  0.732 $\pm$ 0.131 &  0.632 $\pm$ 0.213 &  0.724 $\pm$ 0.162 &  0.664 $\pm$ 0.147 \\
Decision Tree \cite{breiman2017classification}   &  0.613 $\pm$ 0.083 &  0.715 $\pm$ 0.171 &  0.548 $\pm$ 0.204 &  0.709 $\pm$ 0.226 &  0.594 $\pm$ 0.126 \\
Gaussian Naive Bayes \cite{chan1982updating}     &  0.512 $\pm$ 0.101 &  0.586 $\pm$ 0.123 &  0.435 $\pm$ 0.098 &  0.603 $\pm$ 0.210 &  0.493 $\pm$ 0.096 \\
GentleBoost \cite{friedman2001greedy}            &  0.710 $\pm$ 0.061 &  0.741 $\pm$ 0.084 &  0.735 $\pm$ 0.024 &  0.682 $\pm$ 0.129 &  0.736 $\pm$ 0.045 \\
K-Nearest Neighbors \cite{fix1985discriminatory} &  0.752 $\pm$ 0.039 &  0.793 $\pm$ 0.070 &  0.743 $\pm$ 0.032 &  0.765 $\pm$ 0.076 &  0.766 $\pm$ 0.041 \\
Logistic Regression \cite{cox1958regression}     &  \textbf{0.780 $\pm$ 0.098} &  \textbf{0.831 $\pm$ 0.142} &  0.780 $\pm$ 0.048 &  \textbf{0.778 $\pm$ 0.211} &  0.799 $\pm$ 0.072 \\
Random Forest \cite{breiman2001random}           &  0.713 $\pm$ 0.035 &  0.755 $\pm$ 0.095 &  0.724 $\pm$ 0.061 &  0.708 $\pm$ 0.139 &  0.733 $\pm$ 0.025 \\
RUSBoost \cite{seiffert2009rusboost}             &  0.710 $\pm$ 0.029 &  0.721 $\pm$ 0.063 &  0.781 $\pm$ 0.067 &  0.627 $\pm$ 0.123 &  0.746 $\pm$ 0.026 \\
$\text{SVM}_\text{C}$ \cite{chang2011libsvm}     &  0.720 $\pm$ 0.050 &  0.740 $\pm$ 0.118 &  0.791 $\pm$ 0.071 &  0.639 $\pm$ 0.182 &  0.756 $\pm$ 0.031 \\
$\text{SVM}_\text{Q}$ \cite{havlivcek2019supervised}&  \textbf{0.780 $\pm$ 0.062} &  0.793 $\pm$ 0.097 &  \textbf{0.817 $\pm$ 0.042} &  0.738 $\pm$ 0.119 &  \textbf{0.803 $\pm$ 0.057} \\
\bottomrule
\end{tabular}
\end{table*}

\begin{table*}[]
    \centering
    \caption{Indian Pines---multiclass classification performance metrics of different state-of-the-art models and the presented quantum kernel SVM. The models are trained on 4 splits containing 100-100-200 train-validation-test data points from classes \textit{corn-mintill} (class $3$), \textit{soybean-notill} (class $10$), \textit{hay-windrowed} (class $8$) and \textit{grass-pasture} (class $5$). Results are for number of features $n=50$, reported as mean ± standard deviation over 4 splits. Best results for each metric are highlighted in bold.}
    \label{tab:Indian_multiclass_comparison_results}
    \begin{tabular}{lccccc}
        \toprule
         Model & Accuracy & Precision (weighted) & Recall (weighted) & Cohen's kappa & F1 (weighted) \\
        \midrule
        AdaBoost \cite{freund1997decision} & 0.574 $\pm$ 0.113 & 0.603 $\pm$ 0.141 & 0.574 $\pm$ 0.113 & 0.412 $\pm$ 0.147 & 0.556 $\pm$ 0.142 \\
        Decision Tree \cite{breiman2017classification} & 0.657 $\pm$ 0.079 & 0.698 $\pm$ 0.099 & 0.657 $\pm$ 0.079 & 0.523 $\pm$ 0.107 & 0.654 $\pm$ 0.078 \\
        Gaussian Naive Bayes \cite{chan1982updating} & 0.581 $\pm$ 0.072 & 0.627 $\pm$ 0.064 & 0.581 $\pm$ 0.072 & 0.414 $\pm$ 0.111 & 0.579 $\pm$ 0.088 \\
        GentleBoost \cite{friedman2001greedy} & 0.721 $\pm$ 0.049 & 0.749 $\pm$ 0.051 & 0.721 $\pm$ 0.049 & 0.612 $\pm$ 0.064 & 0.723 $\pm$ 0.048 \\
        K-Nearest Neighbors \cite{fix1985discriminatory} & 0.815 $\pm$ 0.034 & 0.829 $\pm$ 0.031 & 0.815 $\pm$ 0.034 & 0.744 $\pm$ 0.045 & 0.813 $\pm$ 0.035 \\
        Logistic Regression \cite{cox1958regression} & 0.740 $\pm$ 0.090 & 0.755 $\pm$ 0.090 & 0.740 $\pm$ 0.090 & 0.638 $\pm$ 0.124 & 0.737 $\pm$ 0.093 \\
        Random Forest \cite{breiman2001random} & 0.723 $\pm$ 0.060 & 0.752 $\pm$ 0.065 & 0.723 $\pm$ 0.060 & 0.615 $\pm$ 0.088 & 0.721 $\pm$ 0.063 \\
        RUSBoost \cite{seiffert2009rusboost} & 0.516 $\pm$ 0.032 & 0.463 $\pm$ 0.055 & 0.516 $\pm$ 0.032 & 0.343 $\pm$ 0.045 & 0.435 $\pm$ 0.018 \\
        $\text{SVM}_\text{C}$ \cite{chang2011libsvm} & 0.826 $\pm$ 0.071 & \textbf{0.854} $\pm$ \textbf{0.042} & 0.826 $\pm$ 0.071 & 0.759 $\pm$ 0.099 & 0.825 $\pm$ 0.072 \\
        $\text{SVM}_\text{Q}$ \cite{havlivcek2019supervised} & \textbf{0.833} $\pm$ \textbf{0.031} & 0.845 $\pm$ 0.032 & \textbf{0.833} $\pm$ \textbf{0.031} & \textbf{0.768} $\pm$ \textbf{0.042} & \textbf{0.830} $\pm$ \textbf{0.033} \\
        \bottomrule
        \end{tabular}
\end{table*}

~

\begin{figure*}[htbp]
    \centering
    \setlength{\fboxsep}{0pt}  
    \setlength{\fboxrule}{1pt} 
    \begin{subfigure}[b]{0.15\textwidth}
        \centering
        \fbox{\includegraphics[width=\dimexpr\textwidth-2\fboxrule\relax]{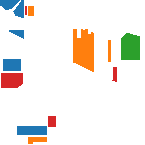}}
        \caption{Ground Truth}
    \end{subfigure}
    \hfill
    \begin{subfigure}[b]{0.15\textwidth}
        \centering
        \fbox{\includegraphics[width=\dimexpr\textwidth-2\fboxrule\relax]{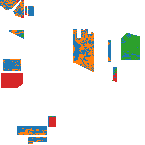}}
        \caption{AdaBoost}
    \end{subfigure}
    \hfill
    \begin{subfigure}[b]{0.15\textwidth}
        \centering
        \fbox{\includegraphics[width=\dimexpr\textwidth-2\fboxrule\relax]{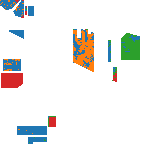}}
        \caption{DT}
    \end{subfigure}
    \hfill
    \begin{subfigure}[b]{0.15\textwidth}
        \centering
        \fbox{\includegraphics[width=\dimexpr\textwidth-2\fboxrule\relax]{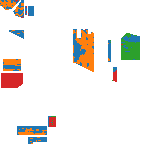}}
        \caption{GNB}
    \end{subfigure}
    \hfill
    \begin{subfigure}[b]{0.15\textwidth}
        \centering
        \fbox{\includegraphics[width=\dimexpr\textwidth-2\fboxrule\relax]{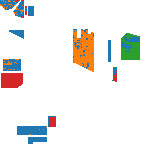}}
        \caption{GentleBoost}
    \end{subfigure}
    \hfill
    \begin{subfigure}[b]{0.15\textwidth}
        \centering
        \fbox{\includegraphics[width=\dimexpr\textwidth-2\fboxrule\relax]{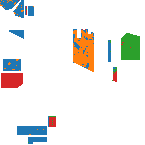}}
        \caption{KNN}
    \end{subfigure}
    
    \vspace{0.5em}
    
    \hspace*{\fill}
    \begin{subfigure}[b]{0.15\textwidth}
        \centering
        \fbox{\includegraphics[width=\dimexpr\textwidth-2\fboxrule\relax]{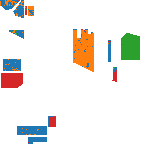}}
        \caption{LR}
    \end{subfigure}
    \hfill
    \begin{subfigure}[b]{0.15\textwidth}
        \centering
        \fbox{\includegraphics[width=\dimexpr\textwidth-2\fboxrule\relax]{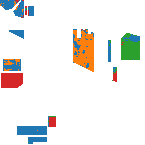}}
        \caption{RF}
    \end{subfigure}
    \hfill
    \begin{subfigure}[b]{0.15\textwidth}
        \centering
        \fbox{\includegraphics[width=\dimexpr\textwidth-2\fboxrule\relax]{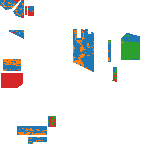}}
        \caption{RUSBoost}
    \end{subfigure}
    \hfill
    \begin{subfigure}[b]{0.15\textwidth}
        \centering
        \fbox{\includegraphics[width=\dimexpr\textwidth-2\fboxrule\relax]{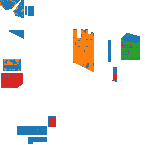}}
        \caption{SVM\_C}
    \end{subfigure}
    \hfill
    \begin{subfigure}[b]{0.15\textwidth}
        \centering
        \fbox{\includegraphics[width=\dimexpr\textwidth-2\fboxrule\relax]{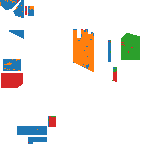}}
        \caption{SVM\_Q}
    \end{subfigure}
    \hspace*{\fill}
    
    \caption{Indian Pines---ground truth and prediction results of different state-of-the-art models and the presented quantum kernel SVM. The models are trained on one split containing 100-100-2763 train-validation-test data points from classes \textit{corn-mintill} (class $3$), \textit{soybean-notill} (class $10$), \textit{hay-windrowed} (class $8$) and \textit{grass-pasture} (class $5$). Results are for number of features $n=50$.}
    \label{fig:wholemap_predictions}
\end{figure*}

\section{Discussion}
\label{sec:discussion}

In this section, we discuss the results presented in Section~\ref{sec:results} in a broader methodological and application-oriented context. Beyond dataset-specific performance, we aim to interpret the observed behavior of the proposed quantum kernel methods in relation to the intrinsic characteristics of hyperspectral data. To this end, we first provide a conceptual discussion on the relevance and limitations of quantum kernel approaches for hyperspectral image analysis, before analyzing the experimental results obtained on the Indian Pines and Methane Detection datasets.

\subsection{Relevance of Quantum Kernel Methods for Hyperspectral Data}
Hyperspectral imagery is characterized by extremely high spectral dimensionality, often comprising hundreds of contiguous bands per pixel, strong inter-band correlations, and subtle class distinctions that are not easily captured by linear models. These properties place hyperspectral data at the intersection of two regimes that are particularly challenging for classical machine learning: high-dimensional feature spaces and limited availability of labeled samples. From the perspective of quantum machine learning, this combination is especially relevant, as the number of spectral features naturally maps to quantum circuit sizes that lie in the many-qubit regime where nontrivial, entangled feature embeddings become possible. In this sense, hyperspectral data provides not only a practically relevant application domain, but also a natural scale at which quantum kernel methods may eventually exhibit advantages that are difficult to reproduce classically.
At the same time, the present study intentionally focuses on quantum embedding circuits that remain classically simulable, despite operating at feature dimensionalities corresponding to hundreds of qubits. This apparent tension is deliberate. By exploring quantum kernels that are computationally tractable yet nontrivial in their entanglement structure, we are able to systematically investigate how such models behave in realistic high-dimensional settings. The goal is not to claim immediate quantum advantage, but rather to develop intuition and design principles for quantum kernel constructions that will become relevant as quantum hardware matures. In this respect, hyperspectral data serves as a valuable testbed: it pushes quantum kernels into regimes of size and complexity that are directly motivated by real sensing modalities, while still allowing controlled experimentation and analysis on classical hardware.

A further intrinsic characteristic of hyperspectral classification is that many classes are defined primarily through spectral similarity rather than hierarchical or compositional structure. Examples include discriminating vegetation species, identifying minerals with overlapping absorption features, or separating crop types with highly correlated reflectance profiles~\cite{Tejasree2024,Hennessy2020}. In such cases, decision boundaries are often driven by nuanced spectral differences across many bands, rather than by spatial patterns or semantic hierarchies. Kernel methods, which operate by constructing similarity measures between samples, are therefore a natural modeling choice. Quantum kernels based on state overlaps extend this paradigm by defining similarity through inner products in a high-dimensional Hilbert space, providing a structurally different notion of spectral similarity than classical distance-based kernels. This makes overlap-based quantum kernels particularly well aligned with the similarity-driven nature of many hyperspectral classification tasks.

An important consideration in this context is the expressive power of the induced feature space. Quantum kernels grant access to highly expressive embeddings, which in principle can enhance separability in complex, high-dimensional data. However, for hyperspectral imagery—where data are structured, correlated, and typically scarce in labeled form—unrestricted expressivity can be counterproductive and lead to poor generalization, a phenomenon that is well documented in hyperspectral image classification, particularly in the small-sample regime and high-dimensional settings references~\cite{Ahmad2022, Hong2021}. Accordingly, this work places particular emphasis on controlling the effective expressivity of quantum kernels through bandwidth optimization. Rather than seeking maximally expressive embeddings, we focus on structured feature spaces that better reflect the smooth and correlated nature of hyperspectral signatures. This balance between expressivity and inductive bias is a central theme in kernel theory and proves especially relevant for hyperspectral data.

We also acknowledge that spatial context plays a decisive role in many hyperspectral image analysis pipelines, and that modern deep learning approaches excel by jointly exploiting spectral and spatial information~\cite{Li2020, paoletti2019deep}. While such models consistently achieve state-of-the-art performance, they are also known to introduce increased architectural complexity, limited interpretability, and a heightened risk of overfitting in data-scarce hyperspectral scenarios, issues that have been repeatedly emphasized in comparative reviews and widely used open-source benchmarks~\cite{Audebert2019}.
In this work, however, we deliberately restrict ourselves to pixel-wise, spectral-only analysis. This choice is motivated by the desire to isolate and understand the contribution of quantum feature mappings themselves, without confounding effects introduced by classical preprocessing, spatial aggregation, or hybrid architectures. In this sense, quantum kernels provide a complementary approach to representation learning, enabling explicit control over feature space geometry and expressivity~\cite{Schuld_2021} rather than relying on architectural depth and large annotated datasets~\cite{GilFuster2024_MLST}. By analyzing unprocessed spectral data, we aim to clearly attribute observed behaviors—both strengths and limitations—to the properties of the quantum kernels, an attribution that becomes considerably more challenging in heavily hybridized models.

Finally, hyperspectral imaging continues to present numerous scenarios in which labeled data remain scarce, such as emerging sensors, rare materials, or localized environmental phenomena,  specifically methane plume detection~\cite{Thompson2017, 2020_Foote}. In these small-sample regimes, kernel methods offer a complementary alternative to deep learning architectures, relying on explicit regularization and similarity-based inference rather than large numbers of learned parameters. The extensive mathematical framework behind kernel methods further enables a principled analysis of model behavior through quantities such as kernel spectra, alignment, and geometric differences. Quantum kernels enrich this framework by providing feature spaces that are explicit yet non-classical, while hyperspectral data supplies a structured, real-world domain in which these properties can be meaningfully examined beyond synthetic benchmarks.

\subsection{Indian Pines}

In Fig.~\ref{fig:time_indian}, we can notice a clear difference in time complexity for the two tested quantum circuit simulation methods. The standard \texttt{statevector} simulator scales exponentially with the circuit size (i.e., its log-scale plot is linear). Calculating $\mathbf{K}_{train}$ on only one training split and with $30$ features (i.e., simulating $30$ qubits) already takes almost an hour, and $\mathbf{K}_{test}$ almost three hours. As expected, \texttt{cuTensorNet} is able to bring down the time complexity to linear (i.e., its log-scale plot is logarithmic), requiring a few seconds for the same operation.

Fig.~\ref{fig:c_optimal_indian} and Table~\ref{tab:fits_indian} show the kernel bandwidth parameters $c^*$ found by maximizing the classification accuracy of the model on $\mathcal{X}_{val}$. 
In the table, we use the $R^2$ coefficient to assess the scaling laws behind the optimized $c^*$.
The coefficient $R^2$ measures the fraction of the variance in the dependent variable explained by the regression model, with $R^2=1$ indicating a perfect fit and $R^2=0$ corresponding to no explanatory power beyond the mean. In plain terms, it quantifies how well the regression captures the trend in the data relative to random scatter. With appropriate data transformations, the $R^2$ coefficient serves as a tool to assess whether the results follow exponential or power-law scaling.
Overall, the results are inconclusive: the $c^*$ values are rather scattered, and fits are inconsistent, likely due to the small size of $\mathcal{X}_{val}$ and potential underperformance of the Bayesian optimizer. Although the results show some tendency toward power-law scaling---which would be consistent with the observations of~\cite{canatar2022bandwidth}---no definitive trend can be established.

The obtained accuracies on $\mathcal{X}_{train}$ and $\mathcal{X}_{test}$ are shown in Fig.~\ref{fig:ACC_indian}. Observe that the \textit{no bandwidth} approach (i.e., setting $c=1$) quickly leads to strong overfitting in the quantum case. By using the bandwidth parameter optimization, we are able to prevent the quantum model from overfitting. For $n\geq75$, it performs indeed worse than the classical model during the training stage, while achieving advantageous mean accuracy on the test data. 
Visually, the greatest discrepancy between the models for the \textit{bandwidth} approach is observed for $n^*=50$ features, where the quantum model has an advantage. However, the Wilcoxon signed-rank test suggests that with only four data splits, the difference is not statistically significant ($p=0.125$). Moreover, the same test run of all the resultant accuracies also does not point to a statistically significant difference between the performances of the models ($p=0.076$).
This outcome indicates that, while the quantum model shows a trend toward better performance, the limited number of data splits prevents us from establishing statistical significance. Thus, the observed advantage should be interpreted as suggestive rather than conclusive.

Fig.~\ref{fig:kernel_analysis_indian} shows the parameters considered for the kernel analysis. The mean and standard deviation clearly show that optimizing the bandwidth prevents the value concentration problem, which is also present in low-dimensional kernels. Moreover, it also reduces kernel expressibility (i.e., it increases $\epsilon$) with high $n$, which allows generalization, as seen in the accuracy plots.
In the case of the geometric difference between classical and quantum kernels, $g(\kappa_C, \kappa_Q)$, we generally observe similar behavior both in the \textit{no bandwidth} and \textit{bandwidth} scenarios. For small $n$ values, the geometric difference is large, and it decreases to a plateau as $n$ increases. In the \textit{no bandwidth} case, the plateau value is around $\sqrt{N}$ (indicated with a dashed, horizontal line), indicating a potential for quantum advantage, which, however, is not realized here. In the \textit{bandwidth} case, the plateau value lies slightly below the limit $\sqrt{N}$. We note an interesting observation: the closest mean geometric difference to the limit $\sqrt{N}$ for large qubit counts ($n>10$) occurs at $n^* = 50$, which coincides with the best performance of the quantum kernel relative to the classical one.
The measure of kernel alignment between the classical and quantum kernels, $\mathcal{A_C}(\kappa_C, \kappa_Q)$, remains relatively high, indicating that the two kernels are generally similar. However, in the \textit{no bandwidth} case, we observe a decreasing trend in alignment with increasing $n$, which may reflect the exponential concentration of the quantum kernel. In the \textit{bandwidth} case, the quantum kernels on average remain very similar to the classical ones, with the notable exception of $n^* = 50$, where a significant drop in alignment occurs. We believe that this discrepancy allowed the quantum kernel to achieve stronger performance at $n^*$.

Fig.~\ref{fig:indian_spectra} shows the average spectra (with respect to splits) of quantum matrices (\textit{no bandwidth} and \textit{bandwidth}) and classical kernel matrices. To explore the behavior at larger system sizes, we display results for $n=50$ and $n=100$.
In the quantum case of \textit{no bandwidth}, the spectrum decays slowly at $n=50$ and becomes almost flat at $n=100$. Inspection across other qubit numbers confirms that this transition from slow decay to nearly constant spectra is robust. From an inductive bias perspective, this means that the kernel has a slight preference for particular ``directions'' in its functional space: it is highly expressive, capable of representing a wide range of functions, but with a strong tendency to overfit, consistent with observations elsewhere in our analysis.
The classical RBF kernel exhibits a spectrum whose decay rate is largely independent of the number of features and shows a clear exponential tail.
In contrast, the quantum kernel with the optimized bandwidth displays a few dominant eigenvalues, while the remaining tail decays exponentially, more sharply than in the classical case. This indicates that the kernel emphasizes specific directions in its functional space, thereby restricting the set of possible functions that can be represented. Such a bias can promote generalization, especially evident for $n=50$, where the fast decay is particularly pronounced.

In addition to the comparison between classical and quantum kernel SVMs presented above, we further benchmark the two studied approaches against several state-of-the-art methods for pixel-wise classification. The results, reported in Table~\ref{tab:Indian_comparison_results}, correspond to a fixed feature dimensionality of $n = n^* = 50$. All additional classical baseline methods are optimized with respect to their hyperparameters using cross-validation on the training and validation data, ensuring a fair comparison (For details see App.~\ref{app:hyperparameters_optimization}). Classification performance is evaluated across all folds and summarized as mean ± standard deviation for each metric. Overall, the quantum kernel SVM and logistic regression consistently achieve the strongest performance, obtaining either the best or second-best results across all considered metrics.

As for the multiclass classification task, presented in Sect. \ref{sub:datasets}, Table~\ref{tab:Indian_multiclass_comparison_results} includes a similar comparison with state-of-the-art methods. In this case, the performance improvement of the quantum kernel SVM is much clearer, both in terms of mean values and consistency across training splits. For visualization purposes, we generate prediction maps on the whole Indian Pines dataset with the same methods trained on one of the splits. The generated maps, shown in Fig. \ref{fig:wholemap_predictions}, also highlight the improved classification consistency of the tested quantum kernel.

\subsection{Methane Detection}

\begin{figure}[!ht]
    \centering
    \includegraphics[width=0.8\linewidth]{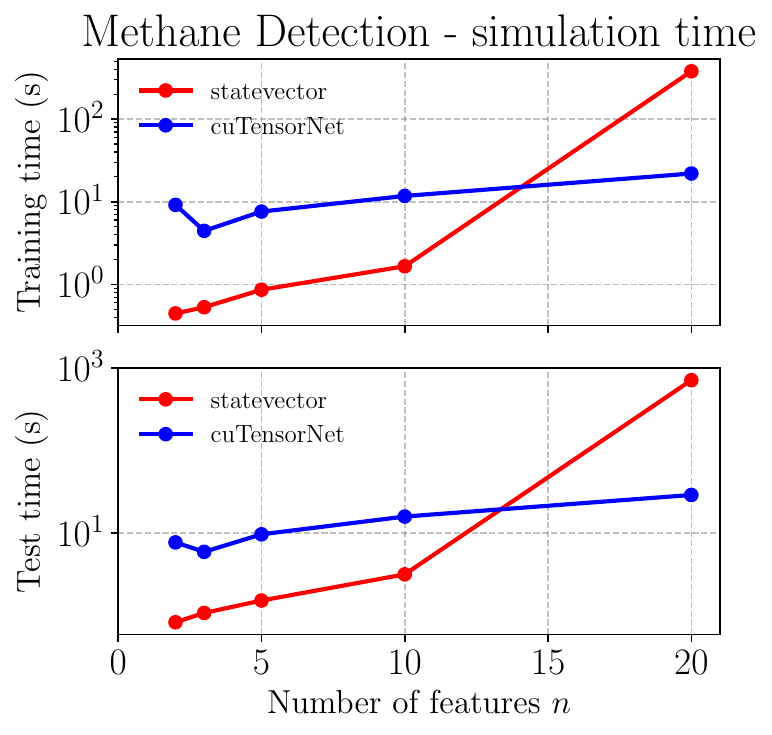}
    \caption{Methane Detection---run time for \texttt{statevector} and \texttt{cuTensorNet} quantum circuit simulation (up to $20$ qubits) for generating $\mathbf{K}_{train}$ and $\mathbf{K}_{test}$ for a Methane Detection training split.}
    \label{fig:time_methane}
\end{figure}

As shown in Fig.~\ref{fig:time_methane}, tensor network contraction for kernel simulation also shows a clear improvement in time complexity with a Methane Detection training split. Having a larger $\mathcal{X}_{train}$ set ($68$ samples instead of $50$), $n=20$ features already make \texttt{statevector} simulation spike the execution time.

\begin{figure}[!ht]
    \centering
    \includegraphics[width=0.9\linewidth]{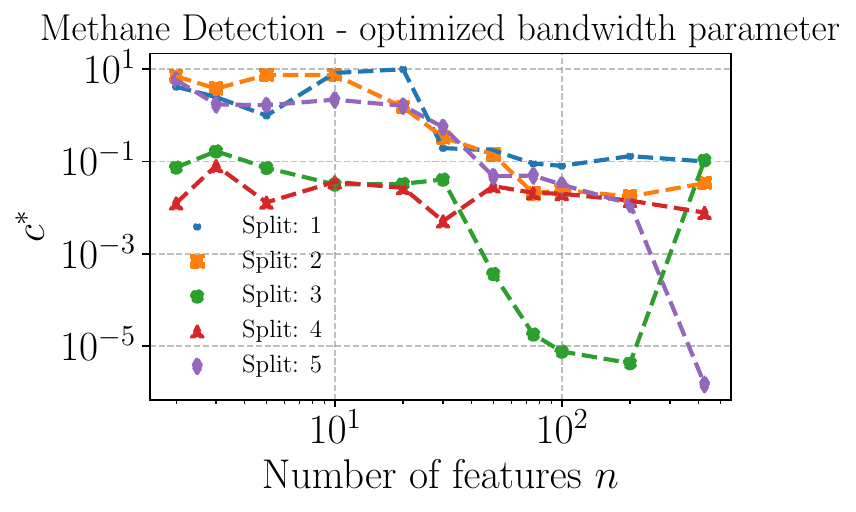}
    \caption{Methane Detection---optimized bandwidth parameter $c^*$ found in the validation stage during the experiment described in the caption of Fig. \ref{fig:ACC_methane} as a function of the number of features $n$. The optimization strategy is described in Section \ref{sub:model_training}. Table \ref{tab:fits_methane} gathers the power-law and exponential fits of the above data. 
    }
    \label{fig:c_optimal_methane}
\end{figure}

The fits of $c^*$, shown in Table~\ref{tab:fits_methane} and Fig.~\ref{fig:c_optimal_methane}, give a similar interpretation as in the Indian Pines case. The power law scaling is preferred; however, the results are inconclusive.

\begin{figure*}[!ht]
    \centering
    \includegraphics[width=\textwidth]{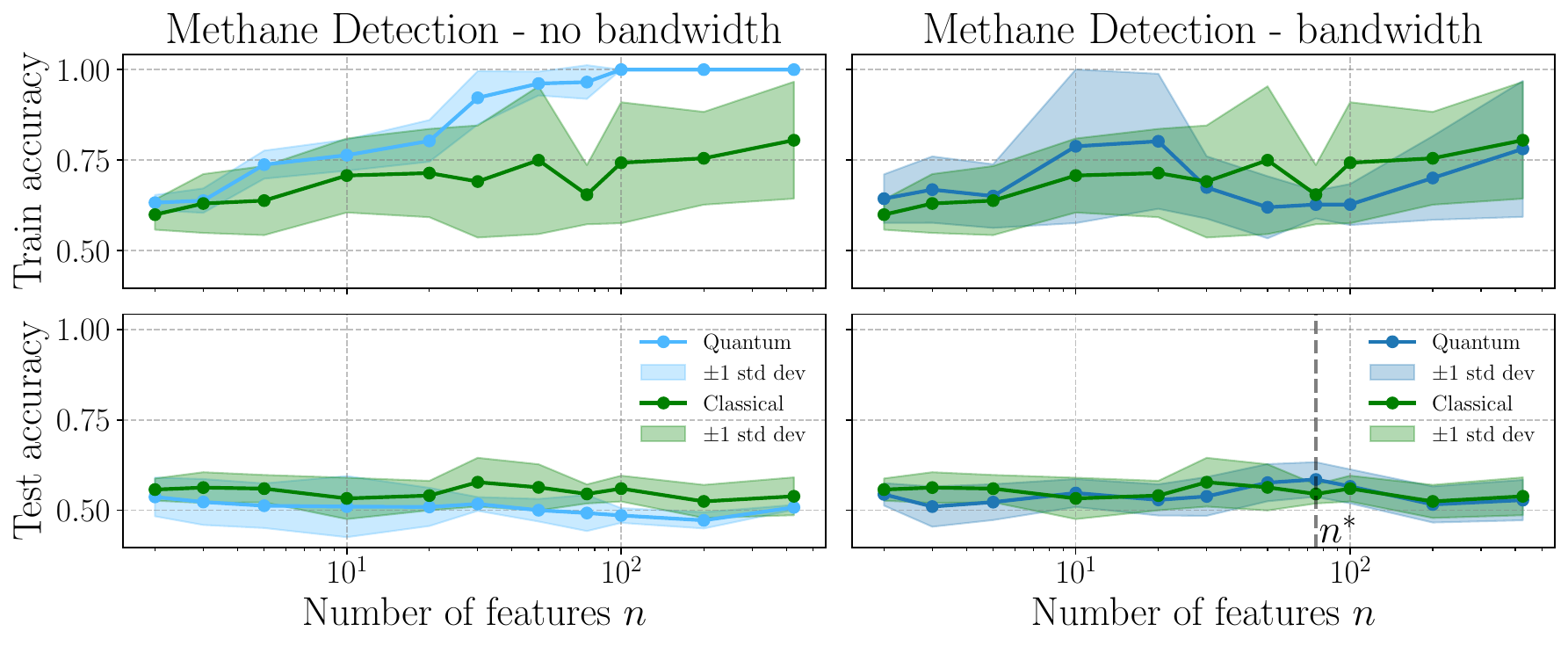}
    \caption{Methane Detection---average train (\textit{top}) and test (\textit{bottom}) accuracy for binary classification of the \textit{background} and \textit{methane} data points from the Methane Detection dataset as a function of the number of features $n$ used. 
    The legend distinguishes between the \textit{quantum} model (described in Section~\ref{sub:qks}) and the \textit{classical} model (described in Section~\ref{sub:model_training}).
    }
    \label{fig:ACC_methane}
\end{figure*}

Fig.~\ref{fig:ACC_methane} shows the average accuracies on the $5$-fold Methane Detection splits. Similarly to the previous experiment, we notice that, in the \textit{no bandwidth} case, the quantum model overfits on the training set with poor (or more precisely, no) generalization capability. The generalization improves when bandwidth optimization is introduced.
Here we notice a discrepancy between the classical and quantum models for $n^*=75$ features, with the quantum model having an advantage. The Wilcoxon signed-rank test for $5$ data splits highlights a higher statistical significance ($p=0.063$) than the $n^*=50$ case of Indian Pines. However, both the classical and quantum models perform poorly on average, making this a challenging classification task, and the same test run of all the resultant accuracies does not point to a statistically significant difference ($p=0.420$).

\begin{figure*}[!ht]
    \centering
    \includegraphics[width=\textwidth]{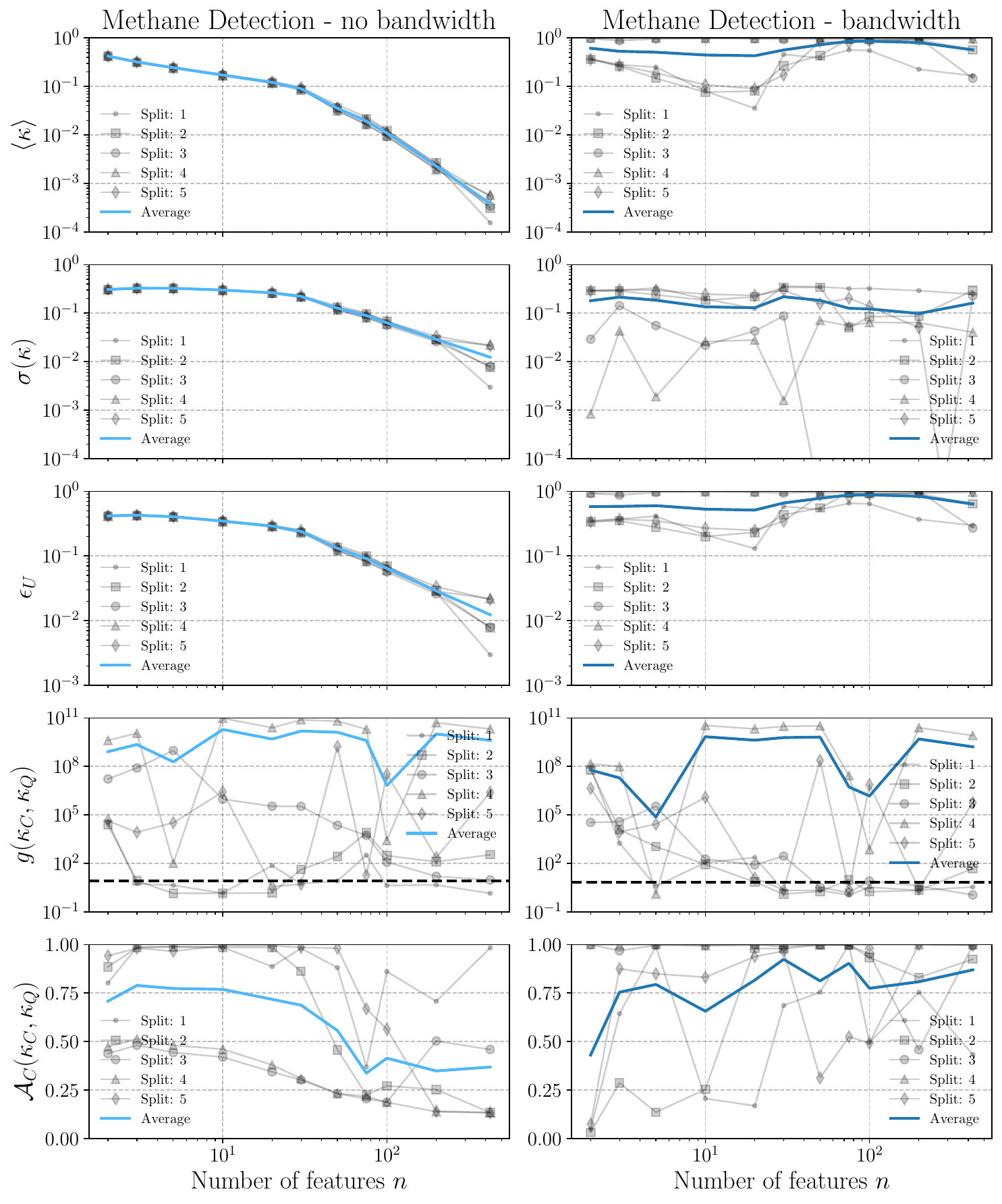}
    \caption{Methane Detection---aggregate plots for the analysis of kernel matrices: mean kernel values $\langle \kappa \rangle$, std of kernel values $\sigma( \kappa)$, expressibility $\epsilon$, geometric difference between classical and quantum matrices $g$, alignment between classical and quantum matrices $\mathcal{A}$.}
    \label{fig:kernel_analysis_methane}
\end{figure*}

Fig.~\ref{fig:kernel_analysis_methane} shows the parameters considered for the kernel analysis. With the mean independent kernel values $\langle \kappa \rangle$, spread $\sigma(\kappa)$, and expressibility $\epsilon_U$ in the comparison between \textit{no bandwidth} and \textit{bandwidth}, conclusions are similar. Introducing bandwidth optimization restricts the expressibility of the quantum embedding unitary and removes the problems with kernel concentration. 
However, in the \textit{no bandwidth} scenario, two distinct trends can be observed: up to around $n=30$ the concentration follows a weak power-law behavior, while beyond $n \approx 30$ the scaling remains power-law but with a noticeably stronger effect. This phenomenon requires further investigation, which lies beyond the scope of this work.
Here, the average geometric difference $g(\kappa_C, \kappa_Q)$ is very high in both quantum scenarios. However, observe that the average is dominated by the extremely big values of geometric difference for split $4$ (and partly split $5$). This is enhanced by a log scale on the $y$-axis. We expect that large values of the geometric difference for the mentioned split(s) are an indication of the hardness of the classification tasks on the data. In the \textit{bandwidth} case for the qubit number above roughly $30$, the geometric difference is close to the limit $\sqrt{N}$. For kernel alignment $\mathcal{A}$, similarly to the Indian Pines dataset, the \textit{no bandwidth} scenario is characterized by a decreasing similarity between the classical and quantum kernels. In the \textit{bandwidth} case, this trend is non-existent, and the alignment between quantum and classical kernels stays high for all qubit numbers.

\begin{figure*}[!ht]
    \centering
    \includegraphics[width=0.9\textwidth]{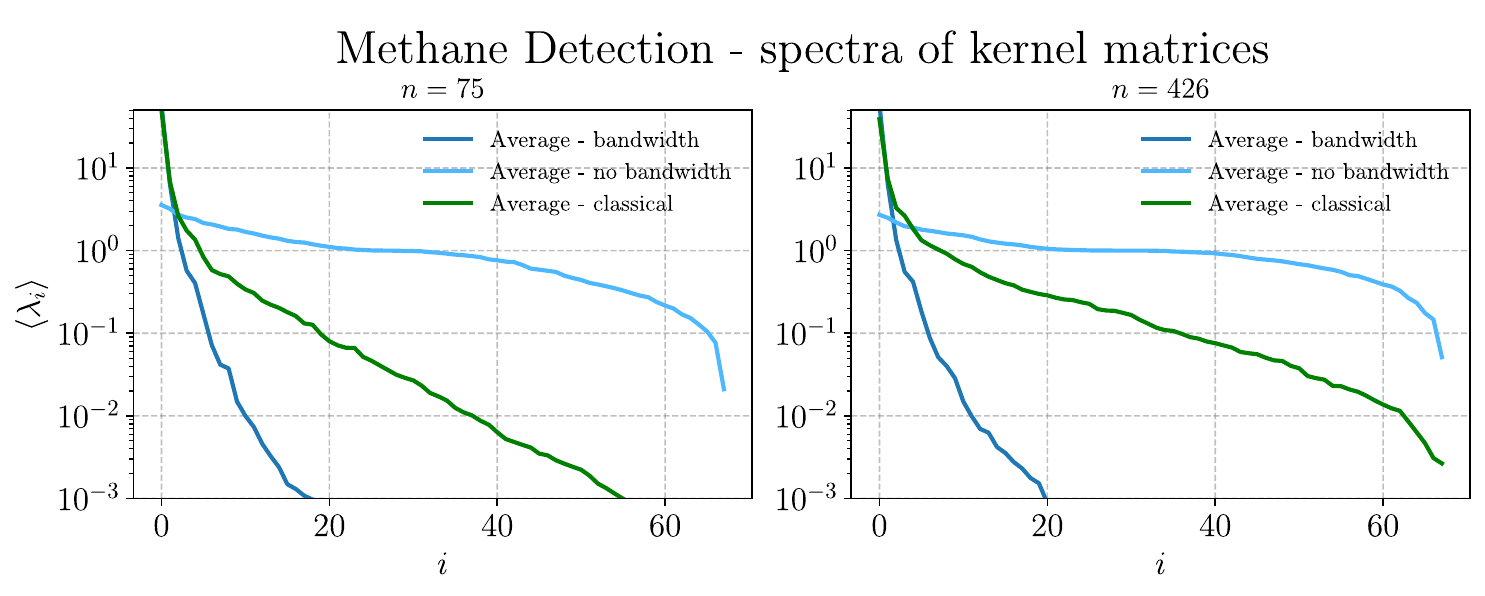}
    \caption{Methane Detection---spectra of the training kernel matrices $\mathbf{K}_{train}$, averaged on all 4 training splits, for selected number of features $n=\{75,426\}$.}
    \label{fig:spectra_methane}
\end{figure*}

Fig.~\ref{fig:spectra_methane} shows the average spectra of the classical kernel and the two quantum kernels (\textit{bandwidth} and \textit{no bandwidth} scenarios). Overall, the eigenvalue behavior is similar to that observed for Indian Pines, with a few notable differences. First, in the \textit{no bandwidth} case, the spectrum is not as flat as for Indian Pines, even though we consider a higher number of features ($n=75$ and $n=426$).
Second, both the classical and bandwidth-optimized quantum kernels exhibit a wider range of eigenvalues: the largest eigenvalues are larger, and the smallest eigenvalues are smaller compared to the Indian Pines case. While the largest eigenvalues of the bandwidth-optimized quantum kernel and the classical kernel are relatively similar, the decay in the quantum case is noticeably sharper.
Despite these differences, the overall interpretation remains consistent: the \textit{no bandwidth} scenario exhibits low inductive bias and is more prone to overfitting, whereas the bandwidth-optimized quantum kernel has much stronger inductive bias. The classical RBF kernel lies between the two quantum cases.

Similar to the Indian Pines dataset, we also compare the SVM-based approaches with several state-of-the-art methods for pixel-wise classification. The results, summarized in Table \ref{tab:Methane_comparison_results}, are obtained using a fixed feature dimensionality of $n = n^* = 75$ (see Fig. \ref{fig:ACC_methane}). All the methods are optimized for each fold with respect to their most important hyperparameters (for the summary, we refer to Appendix A) on the validation data.
Classification performance is assessed across all folds and reported as mean $\pm$ standard deviation for each metric.
Table \ref{tab:Methane_comparison_results} indicates that, in contrast to the Indian Pines case, the Methane detection results are more heterogeneous, making it difficult to identify a single consistently best-performing model. This may suggest a substantially more challenging and complex dataset. Nevertheless, quantum-kernel SVM achieves strong performance, with a significant advantage in the Precision metric.


\begin{table*}[]
    \centering
    \caption{Methane detection performance metrics of small-sample state-of-the-art baseline ML models and the presented quantum kernel SVM. The models are trained on 5 balanced splits. Results are for number of features $n=75$, reported as mean ± standard deviation over 5 splits. Best results for each metric are highlighted in bold.}
    \label{tab:Methane_comparison_results}
\begin{tabular}{llllll}
\toprule
 Model & Accuracy & Precision & Recall & Specificity & $\text{F}_\text{1}$ \\
\midrule
AdaBoost \cite{freund1997decision}  & 0.520 $\pm$ 0.048 & 0.528 $\pm$ 0.093 & 0.467 $\pm$ 0.246 & 0.573 $\pm$ 0.220 & 0.468 $\pm$ 0.140 \\
Decision Tree \cite{breiman2017classification} & 0.535 $\pm$ 0.076 & 0.537 $\pm$ 0.085 & 0.502 $\pm$ 0.120 & 0.569 $\pm$ 0.099 & 0.516 $\pm$ 0.094 \\
Gaussian Naive Bayes \cite{chan1982updating} & 0.563 $\pm$ 0.051 & 0.580 $\pm$ 0.065 & 0.478 $\pm$ 0.100 & \textbf{0.648 $\pm$ 0.103} & 0.519 $\pm$ 0.073 \\
GentleBoost \cite{friedman2001greedy} & 0.558 $\pm$ 0.048 & 0.560 $\pm$ 0.048 & 0.526 $\pm$ 0.088 & 0.590 $\pm$ 0.053 & 0.541 $\pm$ 0.067 \\
K-Nearest Neighbors \cite{fix1985discriminatory} & 0.546 $\pm$ 0.075 & 0.544 $\pm$ 0.074 & 0.458 $\pm$ 0.172 & 0.635 $\pm$ 0.040 & 0.492 $\pm$ 0.127 \\
Logistic Regression \cite{cox1958regression} & \textbf{0.587 $\pm$ 0.050} & 0.583 $\pm$ 0.036 & 0.604 $\pm$ 0.162 & 0.571 $\pm$ 0.100 & \textbf{0.586 $\pm$ 0.093} \\
Random Forest \cite{breiman2001random} & 0.551 $\pm$ 0.033 & 0.557 $\pm$ 0.034 & 0.487 $\pm$ 0.068 & 0.614 $\pm$ 0.046 & 0.519 $\pm$ 0.049 \\
RUSBoost \cite{seiffert2009rusboost} & 0.509 $\pm$ 0.053 & 0.509 $\pm$ 0.085 & 0.456 $\pm$ 0.244 & 0.562 $\pm$ 0.206 & 0.457 $\pm$ 0.142 \\
$\text{SVM}_\text{C}$ \cite{chang2011libsvm} & 0.551 $\pm$ 0.025 & 0.546 $\pm$ 0.028 & \textbf{0.655 $\pm$ 0.229} & 0.447 $\pm$ 0.221 & 0.580 $\pm$ 0.089 \\
$\text{SVM}_\text{Q}$ \cite{havlivcek2019supervised} & 0.585 $\pm$ 0.050 & \textbf{0.601 $\pm$ 0.083} & 0.561 $\pm$ 0.122 & 0.609 $\pm$ 0.154 & 0.571 $\pm$ 0.060 \\
\bottomrule
\end{tabular}
\end{table*}

~

\begin{table}[ht]
\caption{Methane Detection---fit parameters and $R^2$ value of the power law ($log(c^*) = a \cdot log(n) + b$) and exponential ($log(c^*) = a \cdot n + b$) fits to the optimal bandwidth parameters $c^*(n)$. The found bandwidth parameters $c^*$ are presented in Fig. \ref{fig:c_optimal_methane}.
}
\label{tab:fits_methane}
\centering
\begin{tabular}{clrrr}
\toprule
 Split &         Fit &       a &       b &    $R^2$ \\
\midrule
     1 & Power law & $-0.8553$ & $2.2830$ & $0.6019$ \\
    1 & Exponential & $-0.0085$ & $0.1711$ & $0.3199$ \\
    \arrayrulecolor{gray!50}\midrule
    2 & Power law & $-1.3581$ & $3.5338$ & $0.8459$ \\
    2 & Exponential & $-0.0130$ & $0.1427$ & $0.4199$ \\
    \arrayrulecolor{gray!50}\midrule
    3 & Power law & $-1.4195$ & $-0.9855$ & $0.3453$ \\
    3 & Exponential & $-0.0050$ & $-5.2466$ & $0.0235$ \\
    \arrayrulecolor{gray!50}\midrule
    4 & Power law & $-0.1635$ & $-3.4527$ & $0.1409$ \\
    4 & Exponential & $-0.0025$ & $-3.7850$ & $0.1743$ \\
    \arrayrulecolor{gray!50}\midrule
    5 & Power law & $-2.0775$ & $4.6829$ & $0.7139$ \\
    5 & Exponential & $-0.0325$ & $0.5474$ & $0.9433$ \\
\arrayrulecolor{black}\bottomrule
\end{tabular}
\end{table}

\section{Conclusion}


In this work, we demonstrated, to the best of our knowledge, the first large-scale analysis of quantum kernel methods applied to hyperspectral data using the full spectral resolution. This enables the systematic study of quantum kernels in realistic Earth observation scenarios characterized by high dimensionality and limited labeled data.

A central insight of this work is the critical role of bandwidth as a mechanism for controlling the expressivity of quantum kernels in high-dimensional regimes. Properly tuning this parameter enables quantum kernel models to generalize effectively despite operating in exponentially large feature spaces, highlighting the importance of inductive bias in practical quantum machine learning.


Our kernel-level analysis further showed that quantum kernels can exhibit inductive biases that differ substantially from their classical counterparts, and that the effectiveness of these biases is strongly data dependent. In particular, regimes where classical and quantum kernels diverge most strongly emerge as promising candidates for identifying meaningful quantum advantages.


While this work represents an initial step toward systematically evaluating quantum kernels in high-dimensional regimes, many open questions remain. In particular, extending the analysis to more expressive, highly entangled quantum embeddings represents a natural next direction, as such embeddings may unlock richer data representations unavailable to classical kernels. Identifying which types of Earth observation data and problem structures benefit most from these effects, alongside the development of quantum-native explainability tools, remains an important avenue for future research.

\if{False}

\appendices
\section{Work in progress chronology}
\Artur{Let's put here the partial results, the motivation and rationale behind them. In such way we could backtrack all the research directions that we made.}

\Amer{At first, the quantum kernel performed worse. So we introduced data scaling based on kernel bandwidth and optimized its parameters (a, alpha). It performed better. We decided to optimize RBF's parameters, first for all feature numbers together and then for each single feature number, which seems to improve the results.}

\section{Individual fits for the $c^*$'s (Indian Pines)}\label{App:Fits}

In Fig. \ref{fig:c*-fits}, we show individual fits for the $c^*$ parameters for each split.

\begin{figure*}[!ht]
    \centering
    \includegraphics[width=\textwidth]{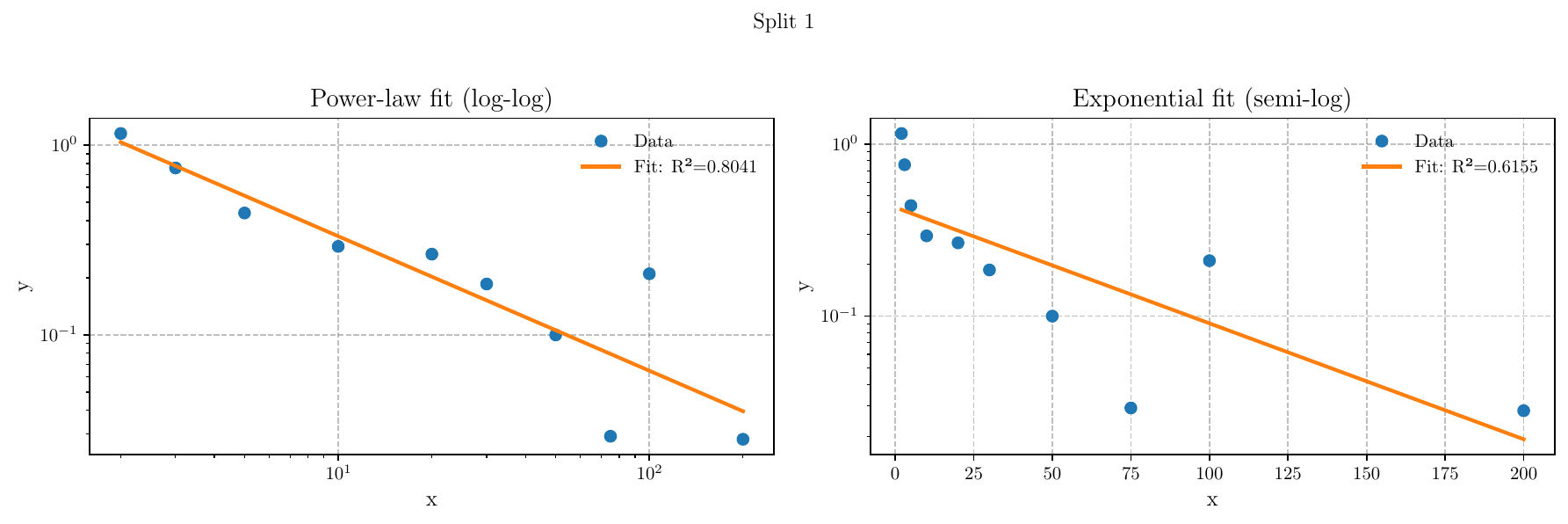}
    \includegraphics[width=\textwidth]{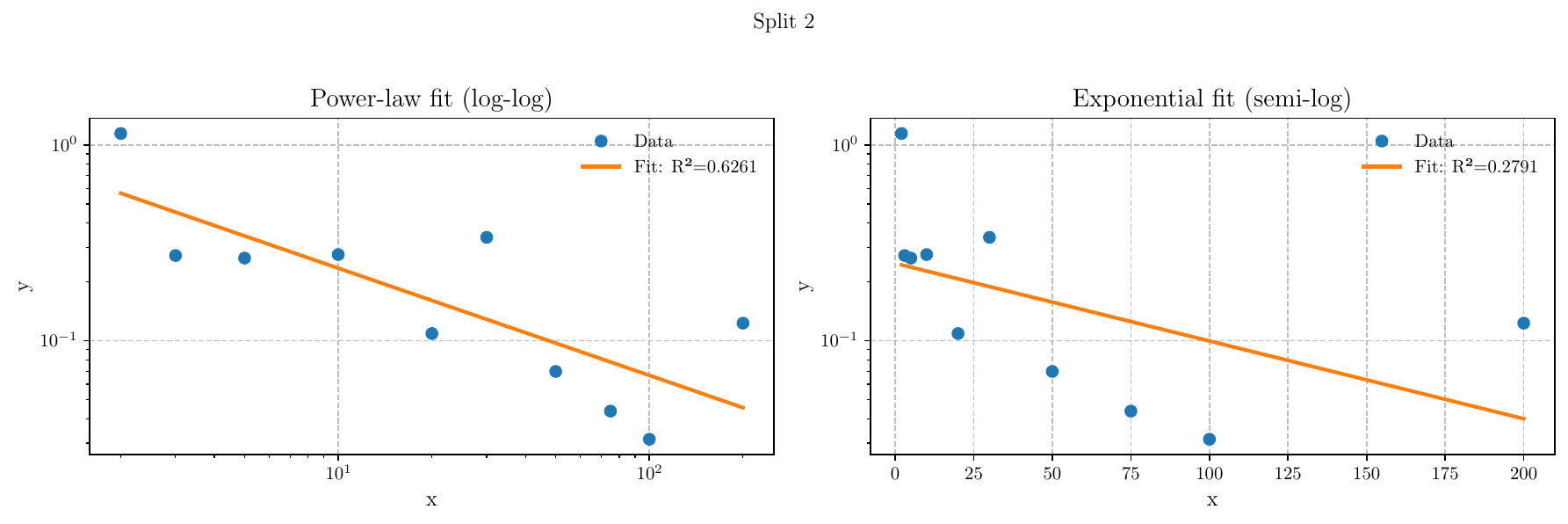}
    \includegraphics[width=\textwidth]{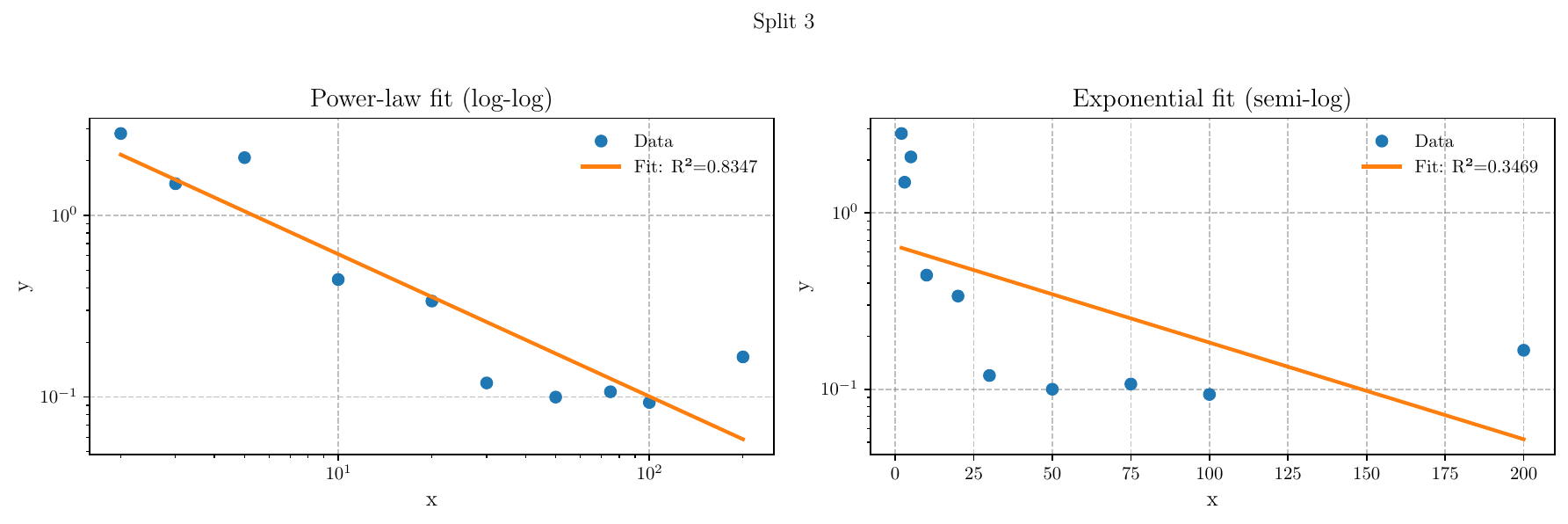}
    \includegraphics[width=\textwidth]{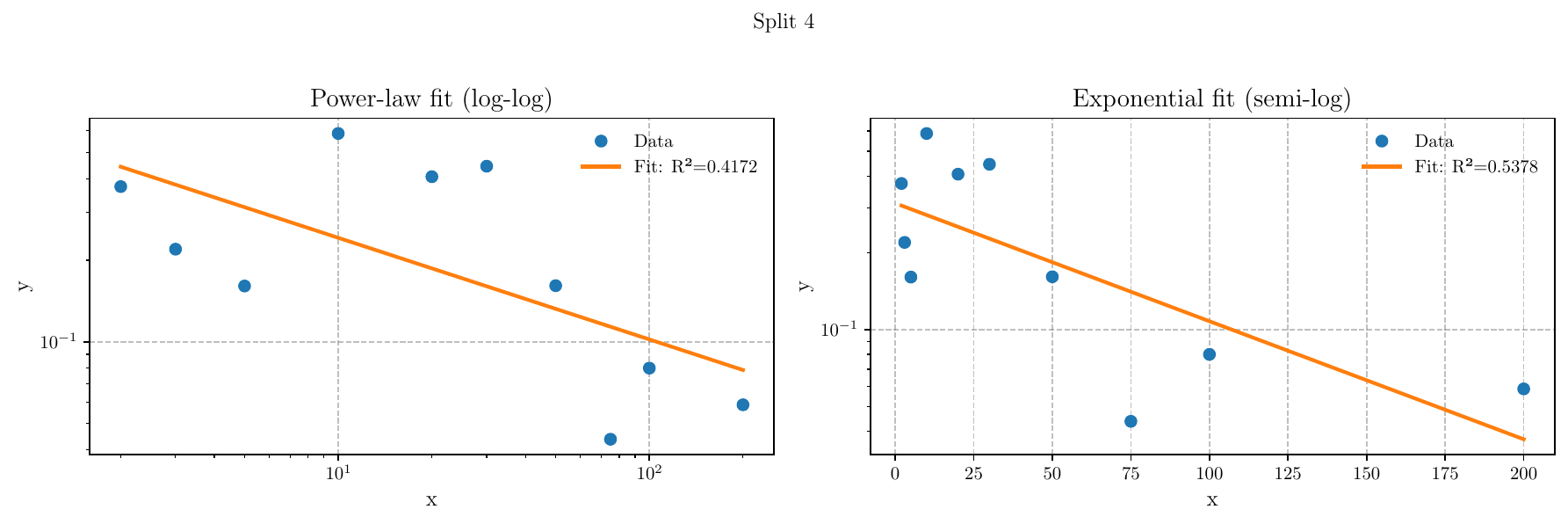}
    \caption{Fits for the optimal bandwidth parameters $c^*$ in the Indian Pines experiment.}
    \label{fig:c*-fits_indian}
\end{figure*}

\fi

\section*{Acknowledgment}

The authors gratefully acknowledge support from the project JUNIQ that has received funding from the German Federal Ministry of Education and Research (BMBF) and the Ministry of Culture and Science of the State of North Rhine-Westphalia. This work is part of the Quantum Computing for Earth Observation (QC4EO) initiative from the ESA $\Phi$-lab, OSIP contract number 4000136455/21/NL/GLC/my. This work is partially supported by the Icelandic Centre for Research, project \textit{"Hybrid Quantum-Classical Workflows for EO"}, RANNÍS grant number: 2511078-051 (\url{http://en.rannis.is/}). AM was supported by the Priority Research Area Digiworld under the program Excellence Initiative – Research University at the Jagiellonian University in Kraków. AMW and JN were supported by the Silesian University of Technology grant for maintaining and developing research potential, as well as by the European Funds for Silesia 2021-2027 Program co-financed by the Just Transition Fund—project entitled ``Development of the Silesian biomedical engineering potential in the face of the challenges of the digital and green economy (BioMeDiG)'' (project number: FESL.10.25-IZ.01-07G5/23). JN was supported by the Silesian University of Technology Rector's grant: 02/080/RGJ25/0052.

\ifCLASSOPTIONcaptionsoff
  \newpage
\fi

\bibliographystyle{IEEEtran}
\bibliography{ref_all}

\appendices

\section{Hyperparameter optimization across models}\label{app:hyperparameters_optimization}
Hyperparameter optimization was performed with Bayesian optimization using Gaussian Processes (\texttt{skopt.gp\_minimize} routine). The search spaces are presented in Table~\ref{tab:hyperparameters}.
\begin{table}[ht!]
    \centering
    \caption{Hyperparameter search spaces for the models used in this work.}
    \label{tab:hyperparameters}
\begin{tabular}{c c c}
\toprule
Model & Parameters & Range\\
\midrule
   $\text{SVM}$  & \texttt{C} & $10^{0}-10^2$\\
          & $c = a\cdot n^{-\alpha}$ & \\
          & $a$ & $10^{-1}-10^1$\\
          & $\alpha$ & $0-3$\\
   \midrule
AdaBoost & \texttt{n\_estimators} & $50,\dots, 300$ \\
         & \texttt{learning\_rate} & $10^{-3}-10^0$ \\
         \midrule
Decision Tree & \texttt{max\_depth} & $2,\dots,15$\\
              & \texttt{min\_samples\_leaf} & $1,\dots,20$\\
              \midrule
Gaussian Naive Bayes &\texttt{var\_smoothing} & $10^{-12} - 10^{-7}$ \\
\midrule
GentleBoost  &\texttt{n\_estimators} & $50,\dots,300$ \\
                &\texttt{learning\_rate} & $10^{-2}-0.5\cdot10^2$ \\
                \midrule
K-Nearest Neighbors & \texttt{n\_neighbors} & $1,\dots,15$\\
\midrule
Logistic Regression & \texttt{C} & $10^{-3} - 10^{2}$ \\
\midrule
Random Forest  & \texttt{n\_estimators} & $100,\dots,400$\\
              & \texttt{max\_depth} & $3,\dots,15$\\
              \midrule
RUSBoost  &\texttt{n\_estimators} & $50,\dots,300$ \\
          &\texttt{learning\_rate} & $10^{-3}-10^2$ \\
          \bottomrule
\end{tabular}
\end{table}

\begin{IEEEbiography}
[{\includegraphics[width=1in,height=1.25in,clip,keepaspectratio]{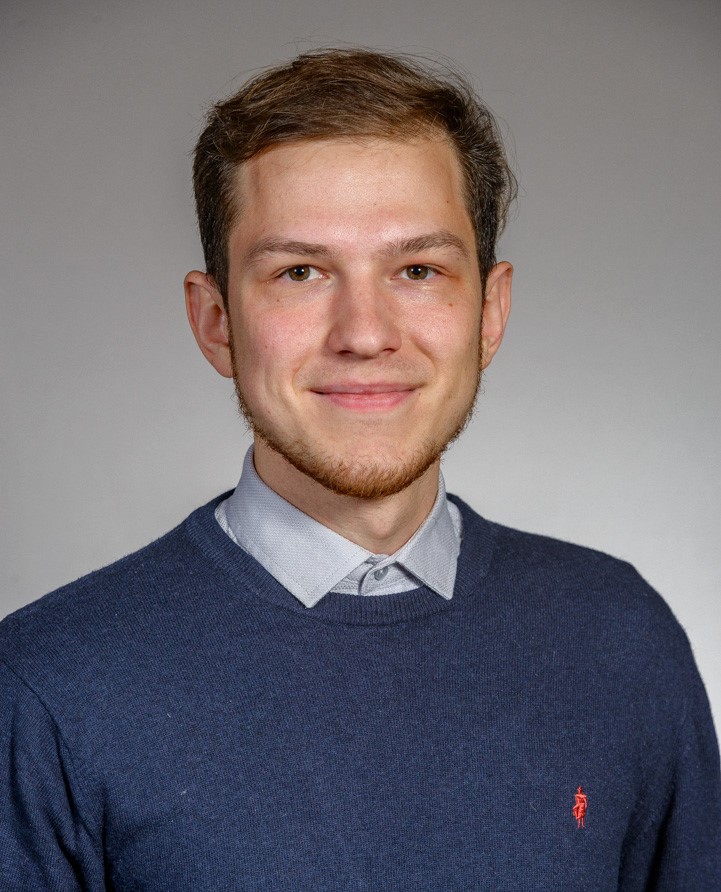}}]{Amer Delilbasic} (Student Member, IEEE) received the B.Sc. and M.Sc. degrees (cum laude) in information and communication engineering from the University of Trento in 2019 and 2021, respectively. He is a member of the ``AI and ML for Remote Sensing'' Simulation and Data Lab at the J\"{u}lich Supercomputing Center, Forschungszentrum J\"{u}lich, Germany. He is currently pursuing the Ph.D. degree in computational engineering at the University of Iceland. His research is mainly focused on machine learning and optimization based on quantum computing and high-performance computing for Earth observation. He has coauthored several articles for reputed journals and conferences for the sector of remote sensing. He has won an ESA OSIP proposal in 2021, and has been a Visiting Researcher with $\mathsf{\Phi}$-lab, European Space Agency ESA/European Space Research Institute ESRIN. He is co-lead of the QC4EO working group within the QUEST IEEE GRSS Technical Committee.
\end{IEEEbiography}

\begin{IEEEbiography}[{\includegraphics[width=1in,height=1.25in,clip,keepaspectratio]{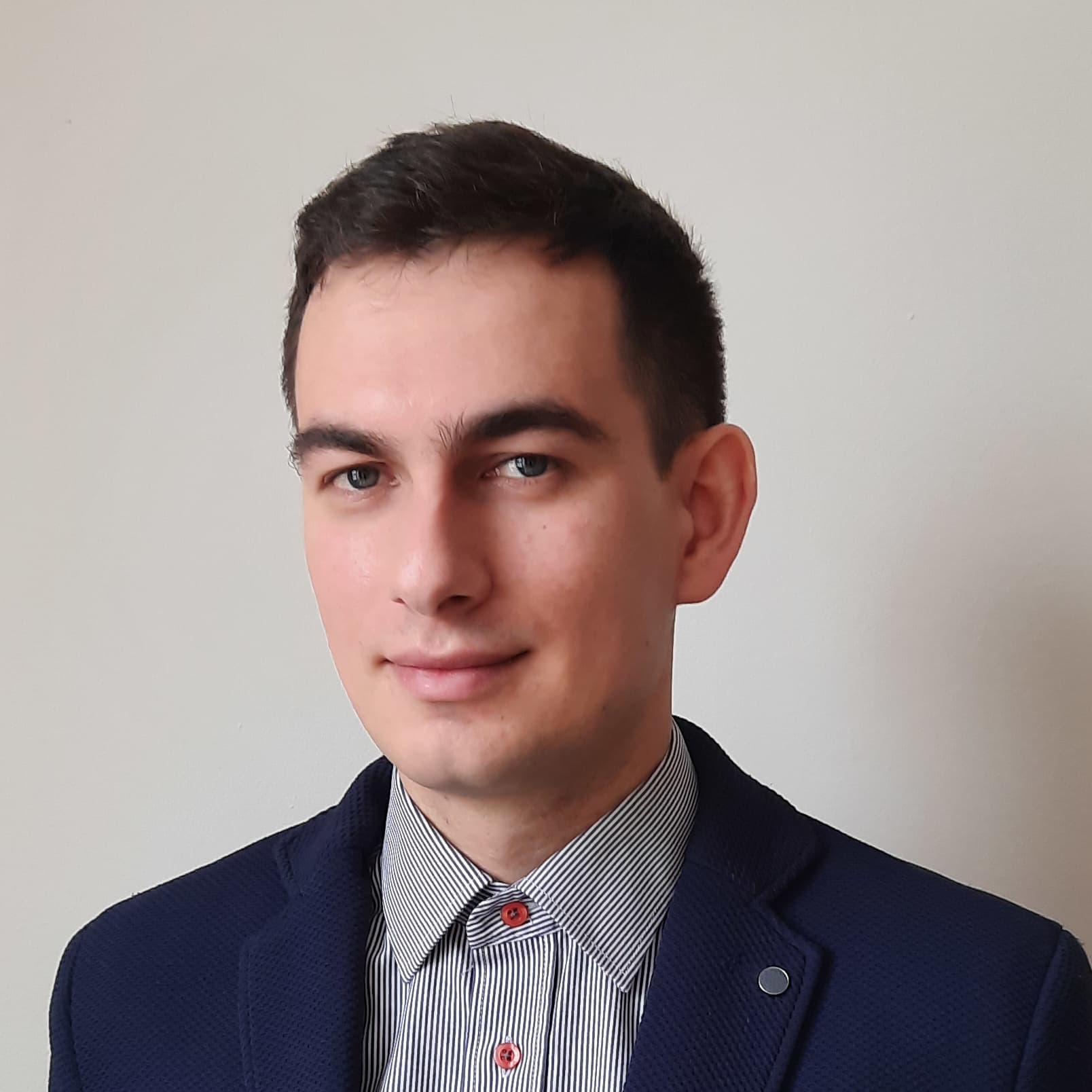}}]{Artur Miroszewski}
received the Ph.D. degree in theoretical physics from the National Centre for Nuclear Research, Otwock, Poland, in 2021. Following his Ph.D., he was affiliated with the Jagiellonian University in Kraków, Poland, where he was involved in European Space Agency projects exploring the potential of quantum machine learning for satellite data analysis. 
He subsequently joined the European Space Agency as a Research Fellow at ESRIN Phi-Lab, where he conducts research in quantum computing and is responsible for quantum computing–related activities in the context of Earth observation. 

He also serves as a quantum computing lecturer at the IEEE GRSS HDCRS summer schools and is the Chair of the QUEST IEEE GRSS Technical Committee.

\end{IEEEbiography}

\begin{IEEEbiography}[{\includegraphics[width=1in,height=1.25in,clip,keepaspectratio]{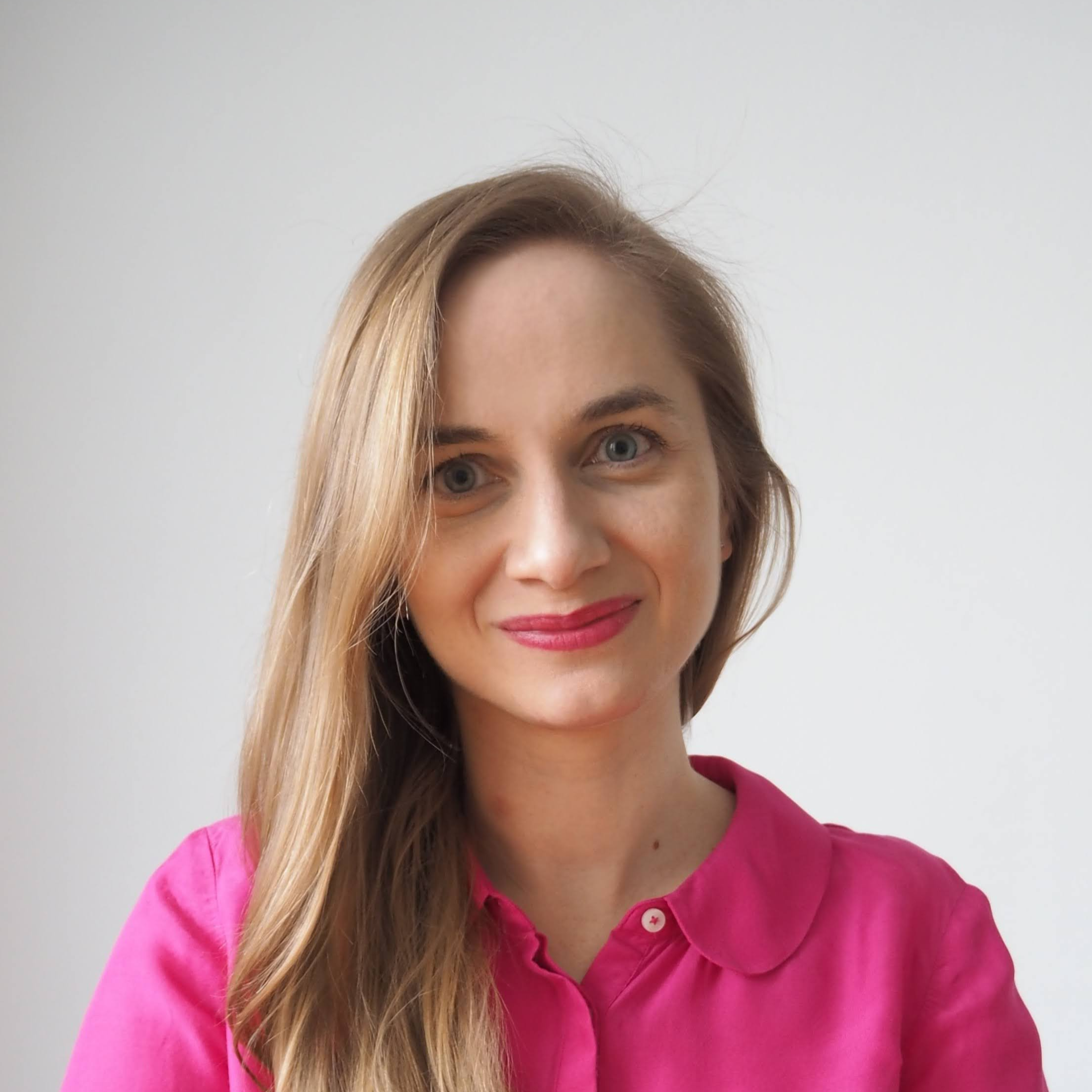}}]{Agata M. Wijata} received her PhD degree in Biomedical Engineering at the Silesian University of Technology, Poland in 2023. Currently, she works as a researcher and Assistant Professor at the Silesian University of Technology (Poland) and as a Machine Learning Researcher at KP Labs (Poland), where she has been focusing on hyperspectral image analysis. Her areas of research interest include multi- and hyperspectral image processing, medical image processing, image-guided navigation systems in medicine, artificial neural networks, and artificial intelligence in general. Agata has been contributing to the Copernicus Hyperspectral Imaging Mission for the Environment---CHIME (European Space Agency) and Intuition-1 (KP Labs) missions from the AI and data processing perspectives. She is a Member of IEEE.
\end{IEEEbiography}

\begin{IEEEbiography}[{\includegraphics[width=1in,height=1.25in,clip,keepaspectratio]{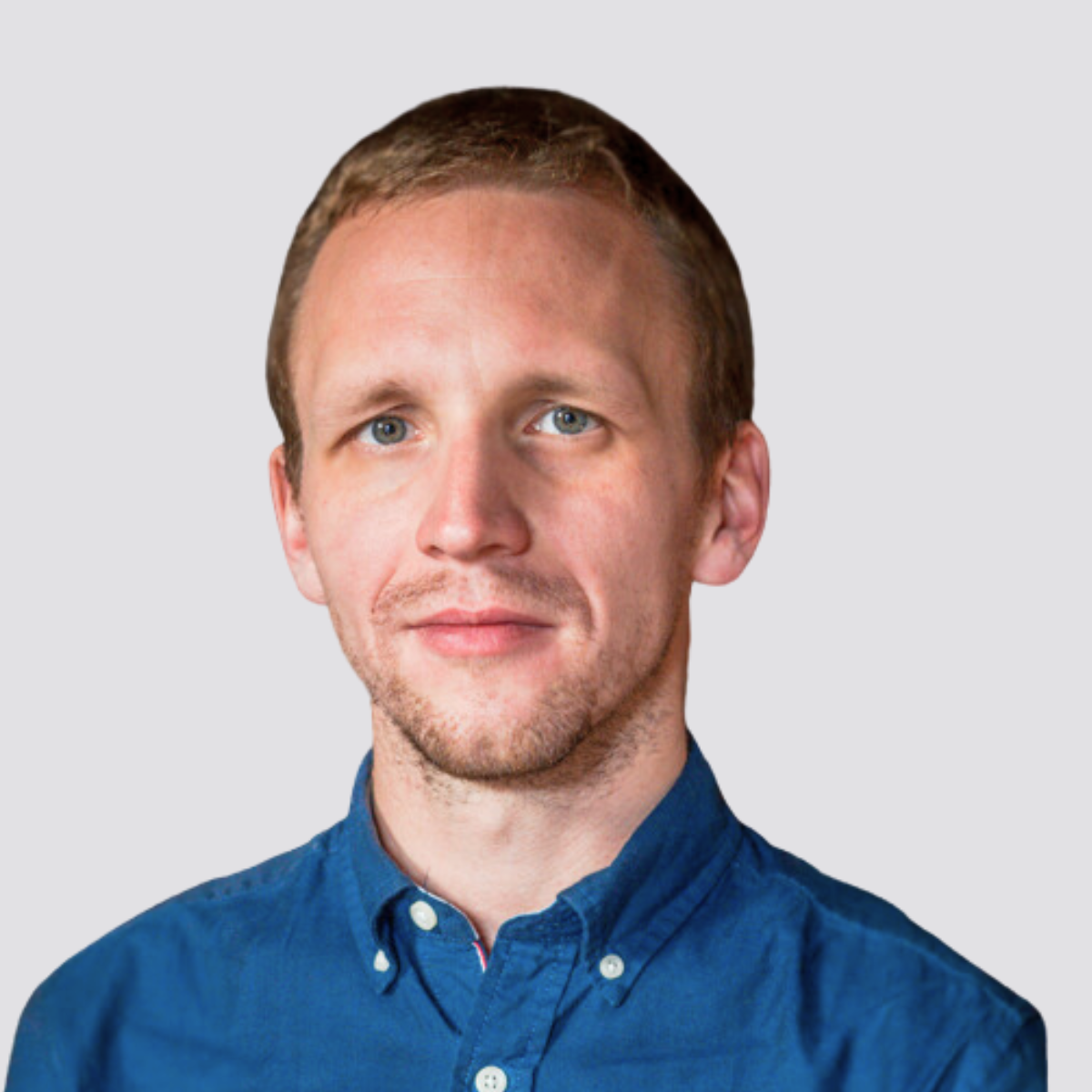}}]{Jakub Nalepa}
received his MSc (2011), PhD (2016), and DSc (2021) in Computer Science from Silesian University of Technology, Poland, where he is an Associate Professor. Jakub is Head of AI at KP Labs where he shapes the scientific and industrial AI objectives of the company related to, among others, EO, on-board and on-the-ground satellite data analysis, machine learning and image analysis. He has been pivotal in designing the on-board deep learning capabilities of the Intuition-1 mission (KP Labs), and has contributed to other missions, including CHIME and OPS-SAT (European Space Agency). His interests focus on (deep) machine learning, HS data analysis, signal processing, remote sensing, and tackling practical challenges which arise in EO to deploy scalable EO solutions. Jakub was the General Chair of the HYPERVIEW Challenge at IEEE ICIP 2022 focusing on the estimation of soil parameters from HSIs on board Intuition-1. He is a Senior Member of IEEE.
\end{IEEEbiography}

\begin{IEEEbiography}[{\includegraphics[width=1in,height=1.25in,clip,keepaspectratio]{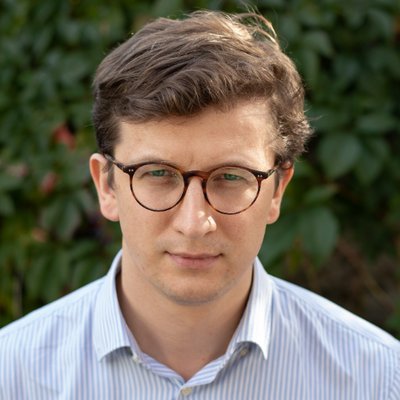}}]{Jakub Mielczarek}
received the dual M.Sc. degrees in theoretical physics and astronomy, the Ph.D. degree in physical sciences, and the habilitation degree
in physical sciences, all from the Jagiellonian 
 University, Krakow, Poland, in 2008, 2012, and 2020,
respectively. He held two postdoctoral fellowships, with the Laboratoire de Physique Subatomique et de Cosmologie (LPSC), Grenoble, France, in 2014–2015
and with the Centre de Physique Theorique (CPT),
Marseille, France, in 2018–2019, accompanied by
numerous shorter internships. He is an Associate Professor with the Jagiellonian University. He is currently leading the Quantum Cosmos Lab, an
interdisciplinary team exploring novel research areas emerging at the interface of theoretical physics and advanced technologies. This notably encompasses
such topics as quantum simulations of gravitational physics and the application of quantum information to space technologies. He is also the Coordinator of
an academic makerspace—Garage of Complexity. This academic unit cultivates a conducive environment for scholars and innovators to transform scientific and technical ideas into reality.
\end{IEEEbiography}

\begin{IEEEbiography}[{\includegraphics[width=1in,height=1.25in,clip,keepaspectratio]{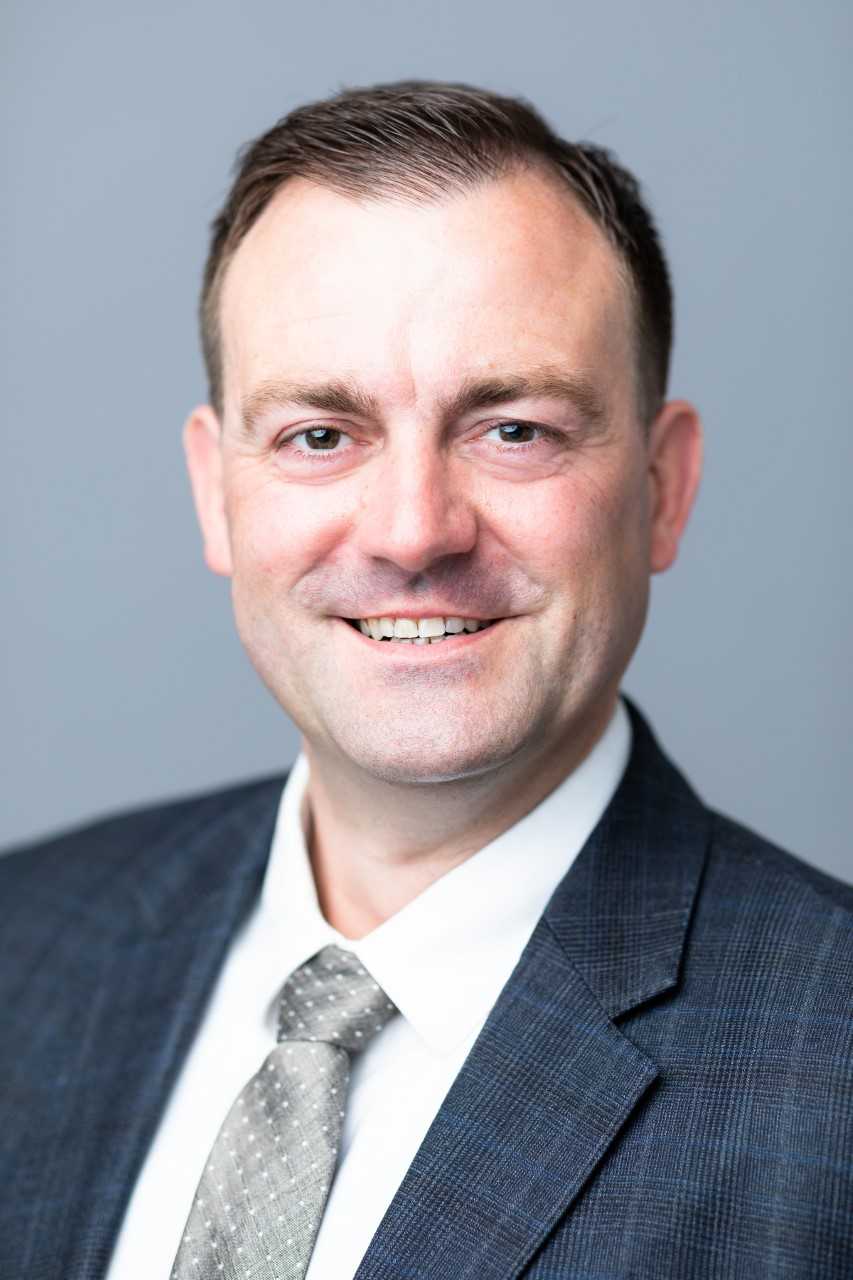}}]{Morris Riedel} (Member, IEEE) received his PhD from the Karlsruhe Institute of Technology (KIT) and worked in data-intensive parallel and distributed systems since 2004. He is currently a Full Professor of High-Performance Computing with an emphasis on Parallel and Scalable Machine Learning at the School of Natural Sciences and Engineering of the University of Iceland. Since 2004, Prof. Dr. - Ing. Morris Riedel held various positions at the Juelich Supercomputing Centre of Forschungszentrum Juelich in Germany. In addition, he is the Head of the joint High Productivity Data Processing research group between the Juelich Supercomputing Centre and the University of Iceland. Since 2020, he is also the EuroHPC Joint Undertaking governing board member for Iceland. His research interests include high-performance computing, remote sensing applications, medicine and health applications, pattern recognition, image processing, and data sciences, and he has authored extensively in those fields. Prof. Dr. – Ing. Morris Riedel online YouTube and university lectures include High-Performance Computing – Advanced Scientific Computing, Cloud Computing and Big Data – Parallel and Scalable Machine and Deep Learning, as well as Statistical Data Mining. In addition, he has performed numerous hands-on training events in parallel and scalable machine and deep learning techniques on cutting-edge HPC systems.
\end{IEEEbiography}

\begin{IEEEbiography}
[{\includegraphics[width=1in,height=1.25in,clip,keepaspectratio]{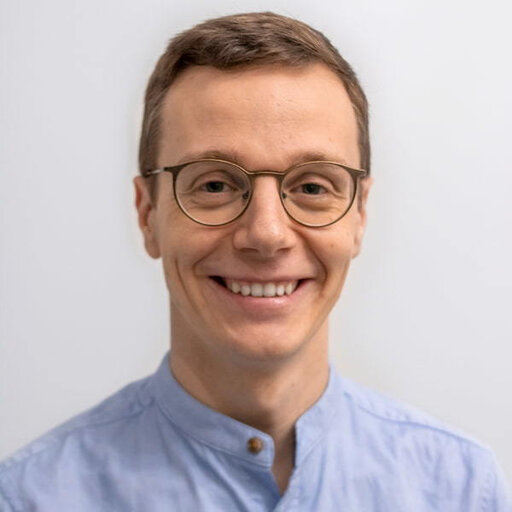}}]{Gabriele Cavallaro} (Senior Member, IEEE) received the B.Sc. and M.Sc. degrees in telecommunications engineering from the University of Trento, Trento, Italy, in 2011 and 2013, respectively, and the Ph.D. degree in electrical and computer engineering from the University of Iceland, Iceland, in 2016. From 2016 to 2021, he has been the Deputy Head of the “High Productivity Data Processing” (HPDP) Research Group, J\"{u}lich Supercomputing Centre, Forschungszentrum Jülich, Germany. Since 2022, he has been the Head of the “AI and ML for Remote Sensing” Simulation and Data Lab at the JSC and an Adjunct Associate Professor with the School of Natural Sciences and Engineering, University of Iceland, Iceland. From 2020 to 2023, he held the position of the Chair for the High-Performance and Disruptive Computing in Remote Sensing Working Group under the IEEE GRSS Earth Science Informatics Technical Committee. In 2023, he took on the role of Co-chair for the ESI TC. Concurrently, he was a Visiting Professor with the $\mathsf{\Phi}$-lab within the European Space Agency (ESA), where he contributes to the Quantum Computing for Earth Observation initiative. In addition, he is an Associate Editor for IEEE TRANSACTIONS ON IMAGE PROCESSING since October 2022. His research interests cover remote sensing data processing with parallel machine learning algorithms that scale on distributed computing systems and innovative computing technologies. Dr. Cavallaro was the recipient of the IEEE GRSS Third Prize in the Student Paper Competition of the IEEE International Geoscience and Remote Sensing Symposium 2015 (Milan—Italy). 
\end{IEEEbiography}





\end{document}